\documentclass[journal]{IEEEtran}
\usepackage[T1]{fontenc}% optional T1 font encoding
\usepackage{cite}
\ifCLASSINFOpdf
\else
\fi

\usepackage{graphicx}
\usepackage[cmex10]{amsmath}
\usepackage{amsthm}
\usepackage{mathrsfs}
\usepackage{txfonts}
\usepackage{stfloats}
\usepackage{cite}
\usepackage{cases}
\usepackage{epstopdf}
\usepackage{inputenc}
\usepackage{longtable}
\usepackage{multirow}
\usepackage{bm}
\usepackage{color,soul}
\usepackage{caption}
\usepackage{subcaption}
\usepackage{algorithm}
\usepackage{algorithmic}

\newtheorem{prop}{Proposition}

\begin{document}

\section*{Disclaimer}
Disclaimer: This work has been accepted for publication in the IEEE Access. Copyright with IEEE. Personal use of this material is permitted. However, permission to reprint/republish this material for advertising or promotional purposes or for creating new collective works for resale or redistribution to servers or lists, or to reuse any copyrighted component of this work in other works must be obtained from the IEEE. This material is presented to ensure timely dissemination of scholarly and technical work. Copyright and all rights therein are retained by authors or by other copyright holders. All persons copying this information are expected to adhere to the terms and constraints invoked by each author's copyright. In most cases, these works may not be reposted without the explicit permission of the copyright holder. For more details, see the IEEE Copyright Policy 

\title{Resource Allocation 
for CoMP in Cellular Networks
with Base Station Switching}

\author{Yoghitha Ramamoorthi
        and Abhinav Kumar,\\
        Department of Electrical Engineering, 
Indian Institute of Technology Hyderabad, Telangana, 502285 India. \\
%502285
Email: \{ee15resch02007, abhinavkumar\}@iith.ac.in }

\maketitle
\begin{abstract}
Base station switching (BSS)
can results in significant reduction in
energy consumption of cellular networks
during low traffic conditions.
We show that the coverage loss due to BSS
can be compensated via
coordinated multi-point (CoMP) based
transmission in a cluster of base stations.
For a BSS with CoMP based system,
we propose various BSS patterns
to achieve suitable trade-offs
between energy savings and 
throughput.
We formulate the CoMP resource allocation
and $\alpha-$Fair user scheduling as a joint
optimization problem.
We derive the optimal time fraction and user scheduling for this problem
and use it to formulate
a simplified BSS with CoMP
optimization problem.
A heuristic that
solves this problem is presented.
Through
extensive simulations,
we show that suitable trade-offs
among energy, coverage,
and rate can be achieved
by appropriately selecting the
BSS pattern,
CoMP cluster, and rate threshold.
\end{abstract}
\begin{IEEEkeywords}
$\alpha$-Fair throughput, base station switching (BSS), cellular networks, coordinated multi-point (CoMP) transmission, downlink, energy.
\end{IEEEkeywords}

\section{Introduction}
%\subsection{Motivation}
The significant increase
in demand of data has led to
deployment of a huge number of
base stations (BSs) in cellular networks.
The BSs consume nearly 80\% of
the total energy consumed in
cellular networks \cite{r1},
out of which 70\% is consumed by power amplifiers, processing circuits,
and air conditioners \cite{r2}.
These BSs are typically
designed and deployed for peak user demands.
However, it has been shown 
in \cite{r4} that the user demand varies
with time resulting in underutilized
BSs and switching off some BSs
during low user demand results in
significant energy savings. 
Further, in \cite{r3},
it has been shown that around 2\% of
global Carbon emission is from cellular networks.
Thus, base station switching (BSS)
during low user demand
is advantageous from both
economical and ecological reasons,
i.e., reduction in energy consumption
and Carbon footprint of the network,
respectively.

In \cite{r4},
a dynamic BSS strategy has been
studied based on the spatial and temporal traces of real-time downlink traffic.
It has been shown in \cite{r6}
that upto 30\% energy can be saved
in a cellular network through BSS.
In \cite{r7},
the energy and throughput trade-offs
for a given coverage have been evaluated.
To overcome the coverage constraint in BSS,
infrastructure sharing through
multi-operator service level agreements
has been proposed in \cite{r8}.
A small cell based approach for BSS
has been presented in \cite{a1} and
\cite{a2}.
Further, a dynamic BSS
strategy based on hybrid
energy supplies has been presented in \cite{bss_bi}.

A promising approach for increasing
edge users performance (equivalently coverage) in cellular networks is coordinated multi-point (CoMP) based
transmission and reception.
Joint transmission (JT), and coordinated scheduling/beamforming are the two variants of CoMP which have been discussed in \cite{rel12}.
In this work,
we consider only CoMP with JT for our analysis
and use CoMP with JT interchangeably with CoMP
throughout the text.
A coverage probability based analysis of
CoMP systems using stochastic geometry
has been derived in \cite{r13}.
Further, in \cite{a4}, it has
been shown through analysis
that CoMP can improve coverage upto
17\%.
The resource allocation for CoMP
has been presented in \cite{r9}.
A new scheduling policy for two tier CoMP network with
one macro-cell and multiple small cells is proposed in \cite{r12}.
However, BSS with CoMP has recently
been studied.

A stochastic geometry based
analysis of outage and coverage probabilities for BSS with CoMP
has been performed in \cite{r14}.
In \cite{r15}, the outage probability for
a hexagonal grid model of BSS with CoMP
in terms of signal-to-noise-ratio (SNR) has been derived.
The energy efficiency analysis of BSS with CoMP,
under the constraint
that only one BS can be switched off,
has been obtained in \cite{r16}.
The fundamental trade-off between energy efficiency and spectral efficiency for BSS with CoMP taking backhaul power consumption into account has been discussed in \cite{r17}.
The performance of BSS with CoMP taking
only uplink into consideration has been
recently investigated in \cite{a3}.
Enlarged coverage and
improved energy savings for
BSS with CoMP has been presented in \cite{green_comp}.
However, the trade-off with respect to
users' throughput has not been
considered in
\cite{green_comp}.
A recent study on JT variant of CoMP has been presented in \cite{comp_jt} that shows
improvements in throughput
at the cost of outage probability.
The trade-off between
energy, coverage, and throughput
for BSS with CoMP has not been jointly
studied in the literature.
Further, suitable resource allocation
schemes for BSS with CoMP
that achieve
these trade-offs are required.
This is the motivation of this work.

The contributions of this paper are
as follows.
\begin{itemize}
\item Various possible
CoMP configurations and BSS patterns
are proposed and compared.
\item The joint BSS and CoMP for cellular networks is formulated as an optimization problem
that is shown to be
a mixed integer non-linear program (MINLP).
\item A decomposed problem of
joint resource allocation and user scheduling for CoMP is formulated as an optimization problem.
Given an $\alpha-$Fair scheduler, optimal user scheduling for CoMP and non-CoMP users, and the optimal resource allocation for a CoMP cluster is derived.
Note that the derived CoMP results
in this work are independent of
the BSs' topology.
\item The optimal CoMP solutions are used to
re-frame a BSS with CoMP optimization problem
that is relatively solvable.
\item A dynamic heuristic is proposed that solves the optimization problem for
an energy efficient point of operation without
compromising on coverage or user rates.
\item The proposed results along with
various CoMP configurations and BSS patterns
are used to achieve the
various trade-offs among energy, coverage,
and throughput.
\end{itemize}

The organization of the paper is as follows.
The system model is described in Section II.
The Joint BSS and CoMP
problem is formulated and analyzed in Section III.
In Section IV, resource allocation and user scheduling for the decomposed CoMP
problem is presented as an optimization problem along with the derivation of the optimal solutions.
The simplified BSS with CoMP optimization problem is re-framed in Section V.
A novel heuristic that solves the
BSS with CoMP problem is described in Section VI.
Extensive numerical results are presented in Section VII.
Some concluding remarks along with possible future works
are discussed in Section VIII.

%\section{CoMP Problem Formulation}
% author names and affiliations
% use a multiple column layout for up to three different
% affiliations

% conference papers do not typically use \thanks and this command
% is locked out in conference mode. If really needed, such as for
% the acknowledgment of grants, issue a \IEEEoverridecommandlockouts
% after \documentclass

% for over three affiliations, or if they all won't fit within the width
% of the page, use this alternative format:
% 
%\author{\IEEEauthorblockN{Michael Shell\IEEEauthorrefmark{1},
%Homer Simpson\IEEEauthorrefmark{2},
%James Kirk\IEEEauthorrefmark{3}, 
%Montgomery Scott\IEEEauthorrefmark{3} and
%Eldon Tyrell\IEEEauthorrefmark{4}}
%\IEEEauthorblockA{\IEEEauthorrefmark{1}School of Electrical and Computer Engineering\\
%Georgia Institute of Technology,
%Atlanta, Georgia 30332--0250\\ Email: see http://www.michaelshell.org/contact.html}
%\IEEEauthorblockA{\IEEEauthorrefmark{2}Twentieth Century Fox, Springfield, USA\\
%Email: homer@thesimpsons.com}
%\IEEEauthorblockA{\IEEEauthorrefmark{3}Starfleet Academy, San Francisco, California 96678-2391\\
%Telephone: (800) 555--1212, Fax: (888) 555--1212}
%\IEEEauthorblockA{\IEEEauthorrefmark{4}Tyrell Inc., 123 Replicant Street, Los Angeles, California 90210--4321}}

% use for special paper notices
%\IEEEspecialpapernotice{(Invited Paper)}

\begin{figure}[t]
%	\begin{center}	
%		\centering
\hspace{-0.2in}
		\includegraphics[height=2.4in,width=\columnwidth]{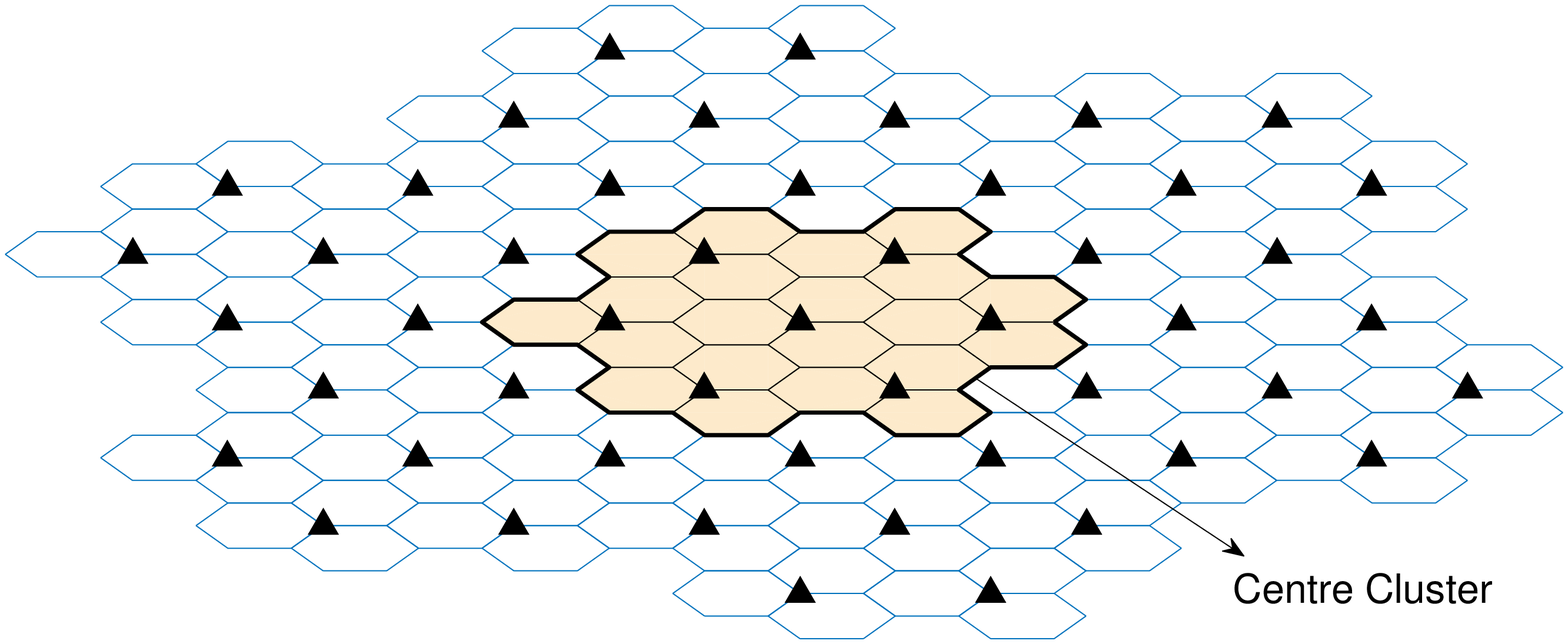}			
		\caption{Benchmark system with the wraparound layout around center cluster (reuse factor 1).}
		\label{fig:1}
%	\end{center}
\end{figure} 

\section{SYSTEM MODEL}

\begin{table}[t]
	\begin{center}
		\caption{Mathematical notations.}	
		\label{tab:mat_not}
		\begin{tabular}{|c|     c      |} 
		    \hline
		    $C_{i}$ & CoMP configurations \\
		    \hline
		    $G_{s}$ & Antenna directivity gain \\ 
		    \hline
		    $h_{u,s}^m$ & Channel gain at user $u$ from sector $s$ on the subchannel $m$ \\
		    \hline
		    $P_{s}^m$ & Power allocated per subchannel $m$ by sector $s$ \\
		    \hline
		    $r_{u,s}$ & link rate of  user $u$ from sector $s$ \\
		    \hline
            $\alpha$ &  Fairness parameter for the $\alpha$-Fair scheduler \\
            \hline
            $\beta_{u,s}$ & Time fraction allocated for user $u$ by a sector $s$\\
            \hline
            $\beta_{u,k}$ & Time fraction allocated for user $u$ by a virtual cluster $k$\\
            \hline
            $\eta(.)$ & Spectral efficiency in $bits/symbol$ \\
            \hline
            $\Gamma_{d}$ & CoMP SINR threshold in $dB$\\            
		    \hline
		    $\gamma_{u,s}^m$ & Received SINR of user $u$ from a sector $s$ \\
		    \hline
		    $\gamma_{u,k}^m$ & Received SINR of user $u$ from a virtual cluster $k$ \\
		    \hline
		    $\lambda_{u}$ & Downlink rate for a user $u$ \\
		    \hline
		    $\mu$ & User density per $km^{2}$\\
			\hline
			$\mathcal{B}$ & Set of BSs with order $B$ \\
             \hline
			$\mathcal{B}_{q}$     & Set of BSs in the cluster
			$q$   \\ %[0.5ex] 			
             \hline
			$\mathcal{E}$  & Percentage energy saved		\\	
			\hline		
			$\mathcal{K}_{q}$ & Set of  virtual clusters in cluster $q$ \\								\hline
			$\mathcal{M}$     & Set of subchannels with order $M$   \\ %[0.5ex] 
			\hline
			$\mathcal{Q}$   &  Set of clusters with order $Q$ \\ 
			\hline
            $R$  & Rate threshold \\
            \hline  
			$\mathcal{S}$ &  Set of sectors with order $S$\\
			\hline
			$\mathcal{S}_{k}$ & Set of sectors in virtual cluster $k$\\
			\hline
			$\mathcal{T}_{\alpha}$ &  $\alpha$-Fair throughput \\
			\hline						
			$\mathcal{U}$  &  Set of users        \\ %[0.5ex]
			             \hline
             
			$\mathcal{U}_{k}$     & Set of users in virtual cluster $k$   \\ %[0.5ex] 	
			\hline
			$\mathcal{V}_{q}$  &  Set of users in the cluster $q$        \\ %[0.5ex]
						\hline
			$\mathcal{W}_{q}$ & Set of sectors in cluster $q$ \\

			\hline
			$\mathcal{Z}_{a1/a2}$ & BSS pattern where $a1$ out of $a2$ BSs are switched off\\
			\hline
			$\mathbb{U}$ & Utility function for $\alpha$-Fair scheduler\\
			\hline
			$|.|$ & Cardinality of a set\\											\hline
			$\lceil . \rceil $ & Ceil the input to smallest following integer\\				
			\hline			
		\end{tabular}
	\end{center}	
\end{table}

% make the title area

\subsection{Benchmark System}
\begin{figure*}[t]
\begin{subfigure} {0.35\textwidth}
\includegraphics[width=0.9\linewidth, height=5cm]{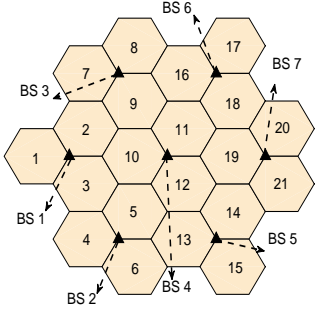} 
\caption{Configuration 1 ($C_{1}$)}
\label{fig:centre_c}
\end{subfigure}
\begin{subfigure}{0.35\textwidth}
\includegraphics[width=0.9\linewidth, height=5cm]{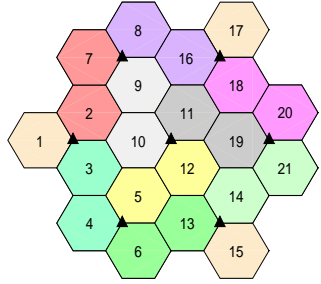}
\caption{Configuration 2 ($C_{2}$)}
\label{fig:sch1}
\end{subfigure}
\begin{subfigure}{0.35\textwidth}
\includegraphics[width=0.9\linewidth, height=5cm]{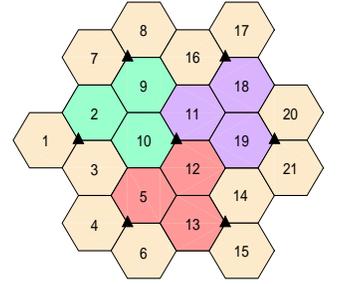}
\caption{Configuration 3 ($C_{3}$)}
\label{fig:sch2}
\end{subfigure}
\caption{Various CoMP configurations for the center cluster.}
\label{fig:c_c}
\end{figure*}

We consider a homogeneous OFDMA
based LTE cellular network as shown in Fig.~\ref{fig:1}.
The set of BSs and corresponding sectors in the network
are denoted by $\mathcal{B}=\{1, 2, ..., B\}$ and $\mathcal{S}=\{1, 2, ..., S\}$,
respectively.
Note that the BSs are represented by triangles in Fig.~\ref{fig:1}.
The hexagons represent the corresponding sectors of a BS such that each BS has three
sectors.
Without any loss of generality,
we assume that the set of sectors is
ordered with the set of BSs.
Hence, any BS $b \in \mathcal{B}$ corresponds
to the sectors $3b-2$, $3b-1$, and $3b$,
in the set $\mathcal{S}$.
For example, in Fig.~\ref{fig:centre_c},
BS 4 corresponds to sectors 10, 11, and 12.
We denote the set of users in the system
by $\mathcal{U}=\{1, 2, ..., U\}$.
We consider that the users are uniformly distributed
in the system for a given user density $\mu$.
Let $\mathcal{M}=\{1, 2, ..., M\}$ denote the set of subchannels available
in the network.
We consider a reuse factor of 1. Hence,
a total of $M$ subchannels are allotted to each sector in $\mathcal{S}$.
A comprehensive list of
mathematical notations used in this paper
is presented in Table~\ref{tab:mat_not}.
Next, we present the channel model considered in this paper.

\begin{table*}
	\begin{center}	
		\caption{Modulation and coding scheme \cite{r18}.}
		\label{tab:1}
		\begin{tabular}{|c|c|c|c|c|c|c|c|c|c|c|c|c|c|c|c|} 
			\hline
			SINR Threshold (dB) & -6.5 & -4 & -2.6 & -1 & 1 & 3 & 6.6 & 10 & 11.4 & 11.8 & 13 & 13.8 & 15.6 & 16.8 & 17.6\\
		%	\hline
%			$\gamma_{i,j}^m(linear)$ & 0.2239 & 0.3981 & 0.5495 & 0.7943 & 1.2589 & 1.9953 & 4.5709 & 10 & 13.8038 & 15.1356 & 19.9526 & 23.9883 & 36.3078 & 147.8630 & 57.5440\\
           % $\gamma_{i,j}^m$ (linear)  & 0.22 & 0.40 & 0.55 & 0.79 & 1.26 & 2.00 & 4.57 & 10 & 13.80 & 15.14 & 19.95 & 23.99 & 36.31 & 47.86 & 57.54\\
			\hline
			 Efficiency  (bits/symbol) & 0.15 & 0.23 & 0.38 & 0.60 & 0.88 & 1.18 & 1.48 & 1.91 & 2.41 & 2.73 & 3.32 & 3.9 & 4.52 & 5.12 & 5.55\\
			\hline
		\end{tabular}
	\end{center} 
\end{table*}

\subsection{Channel Model}
We consider a time division duplex (TDD) system.
For mathematical brevity,
we assume a frequency flat channel model and focus
on the downlink.
However, a similar analysis is possible for
a frequency selective channel and uplink.
The downlink signal-to-interference-plus-noise ratio (SINR)
of a user $u$ from a sector $s$,
denoted by $\gamma_{u,s}^m$,
on a subchannel $m$ is given as
\begin{equation}\label{eq: sinr_dl}
	\gamma_{u,s}^m=\frac{P_{s}^mh_{u,s}^m}{\sum\limits_{\substack {\hat{s} \neq s \\ \hat{s} \in \mathcal{S}}}P_{\hat{s}}^mh_{u,\hat{s}}^m + \sigma^2 } \, ,
\end{equation}
where, $P_{s}^m$ is the power allocated to the subchannel $m$ by
the sector $s$, $\sum\limits_{\substack{\hat{s} \neq s \\ \hat{s} \in \mathcal{S}}}^mP_{s}^mh_{u,\hat{s}}^m $ is the interference
on the subchannel $m$, $\sigma^2$ is the noise power, and $h_{u,s}^m$ denotes
the channel gain between the sector $s$ and the user $u$.
The channel gain is given by
\begin{equation}\label{channel_gain}
	h^{m}_{u,s}=10^{\Bigg(\dfrac{-PL(d)+G_{s}(\phi)+G_{u}-\upsilon-\rho}{10}\Bigg)} \, ,
\end{equation}
where,
$G_{u}$ is the antenna gain,
$\upsilon$ is the penetration loss,
$\rho$ is the loss due to fading and shadowing,
$PL(d)$ is the path loss for the distance $d$ between $u$ and $s$,
and $G_{s}(\phi)$ is the directivity gain equal to
\begin{equation}
	G_{s}(\phi)=25-min \left\{12\left(\frac{\phi}{70}\right)^{2},20\right\}, \forall  -\pi \leq \phi \leq \pi \, ,
\end{equation}
in which $\phi$ denotes the angle between the $u$ and the main lobe
orientation of $s$ \cite{r20}.

\subsection{Resource Allocation and User Scheduling}
Let $P_{BS}$ denote the total transmit power of a BS.
Then,
given that the BS transmit power is shared among the three
sectors of a BS,
the power allocated in a sector $s$ per subchannel $m$, $P_{s}^m$,
is given by
\begin{equation}\label{eq: power per bs}
	P_{s}^m=\frac{P_{BS}}{3\textit{M}} , \ \ \forall \ s \in \mathcal{S}, \ m \in \mathcal{M} \, .
\end{equation}

There exist energy efficient power allocation schemes in the literature \cite{sur}
which save energy through
efficient transmit power allocation.
However,
it has been shown in \cite{sur}
that the energy savings through BSS
is an order of magnitude higher than
energy efficient power allocation schemes.
Thus, in this work,
we consider uniform power allocation
such that the available transmit power per BS is allocated equally among all available subchannels across all the sectors in a BS.
This also corresponds to frequency flat fading.
The presented
analysis can be generalized to frequency selective
fading by using water filling based power allocation schemes as discussed in \cite{WFA}.

We use $\eta(\gamma_{u,s}^m)$ to denote the spectral efficiency achieved by
a user in bits/symbol.
The value of $\eta(\gamma_{u,s}^m)$
obtained from an adaptive modulation and coding
scheme (MCS) is given in Table \ref{tab:1}
for various ranges of SINR \cite{r18}.
Given $\gamma_{u,s}^m$ as in (\ref{eq: sinr_dl}), the
link rate for the user $u$ from sector $s$, denoted by $r_{u,s}$, is expressed as
\begin{equation}\label{eq:l_r}
	r_{u,s} =\frac{\eta(\gamma_{u,s}^m) \, SC_{OFDM} \, SY_{OFDM}}{T_{sc}}M \, ,
\end{equation}
where, $SC_{OFDM}$, $SY_{OFDM}$, and $T_{sc}$ represent
the number of subcarriers per subchannel, number of symbols used per subcarrier,
and time duration of a subframe, respectively.
The factor $M$ represents number of subchannels used in downlink per sector $s$.

We consider an $\alpha-$Fair time based
scheduler at each sector $s$
such that the scheduler allocates all the $M$ subchannels for a downlink time fraction denoted by $\beta_{u,s}$ to a user $u$ associated with it.
In the benchmark system, we assume that
any user $u$ associates with the sector $s$
from which it
receives maximum received SINR on the downlink.
Thus, for a user $u$, $\beta_{u,s}$ is non-zero for only one sector $s$.
The resultant downlink rate for any user $u$,
represented by $\lambda_{u}$,
is given by
\begin{equation}
\lambda_{u} = \sum\limits_{s \in \mathcal{S}}\beta_{u,s}r_{u,s} \, ,
\label{lambda}
\end{equation}
where, $r_{u,s}$ is the link rate as computed in (\ref{eq:l_r}).
The utility function for an $\alpha$-Fair user scheduler 
is expressed as \cite{r19}
\begin{equation}\label{eq:u_alp}
\mathbb{U}_{\alpha}(\lambda)=\begin{cases}
                               \dfrac{\lambda^{1-\alpha}}{1-\alpha}, & \text{$\alpha>0, \ \alpha\neq1$,} \\
                               log(\lambda), & \text{$\alpha=1$.}
                               \end{cases}
\end{equation}
To focus on the downlink,
we consider the TDD downlink time fraction as 1.

\begin{figure*}[t]
\begin{subfigure}{0.35\textwidth}
\includegraphics[width=0.9\linewidth, height=5cm]{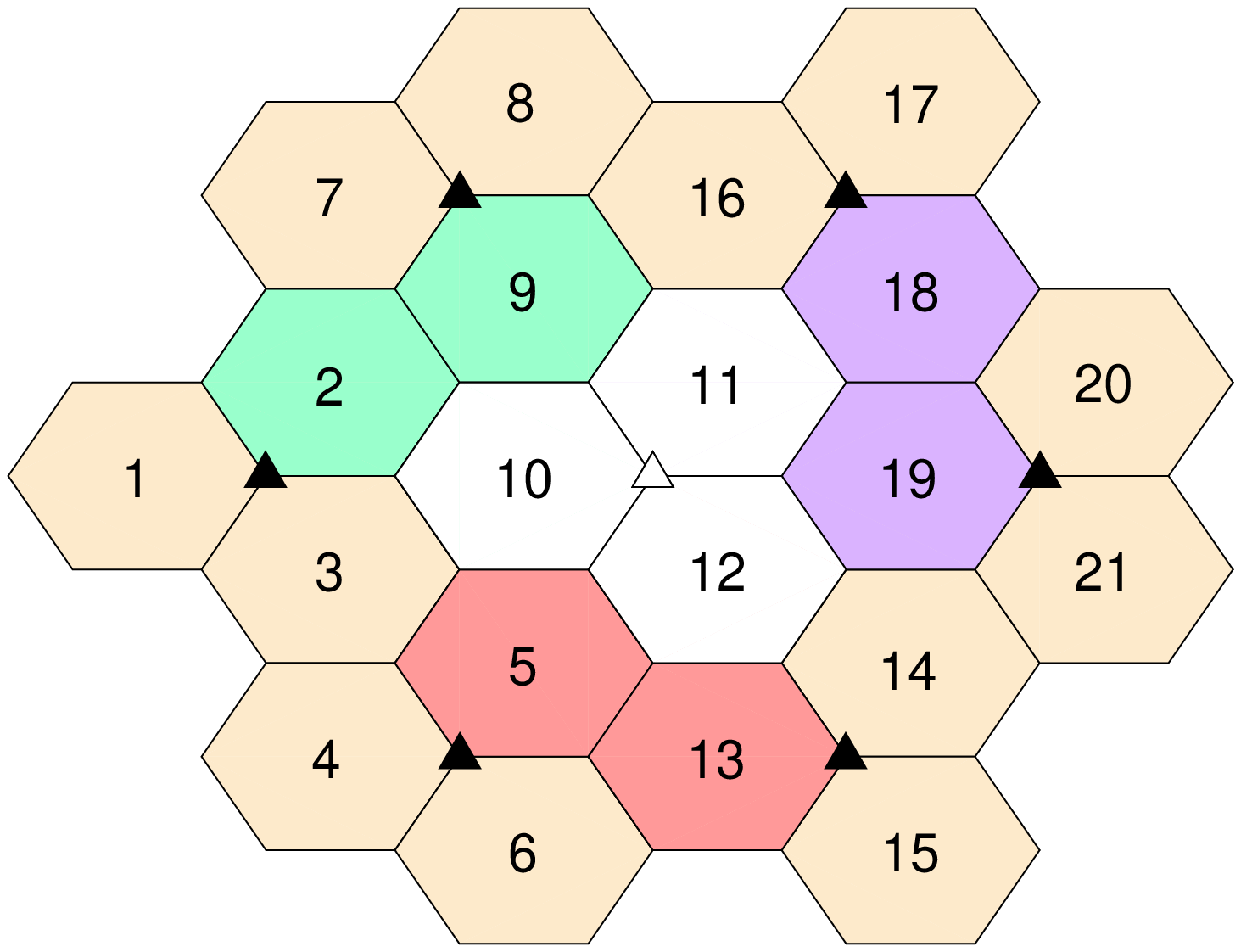} 
\caption{$\mathcal{Z}_{1/7}$}
\label{fig:z1}
\end{subfigure}
\begin{subfigure}{0.35\textwidth}
\includegraphics[width=0.9\linewidth, height=5cm]{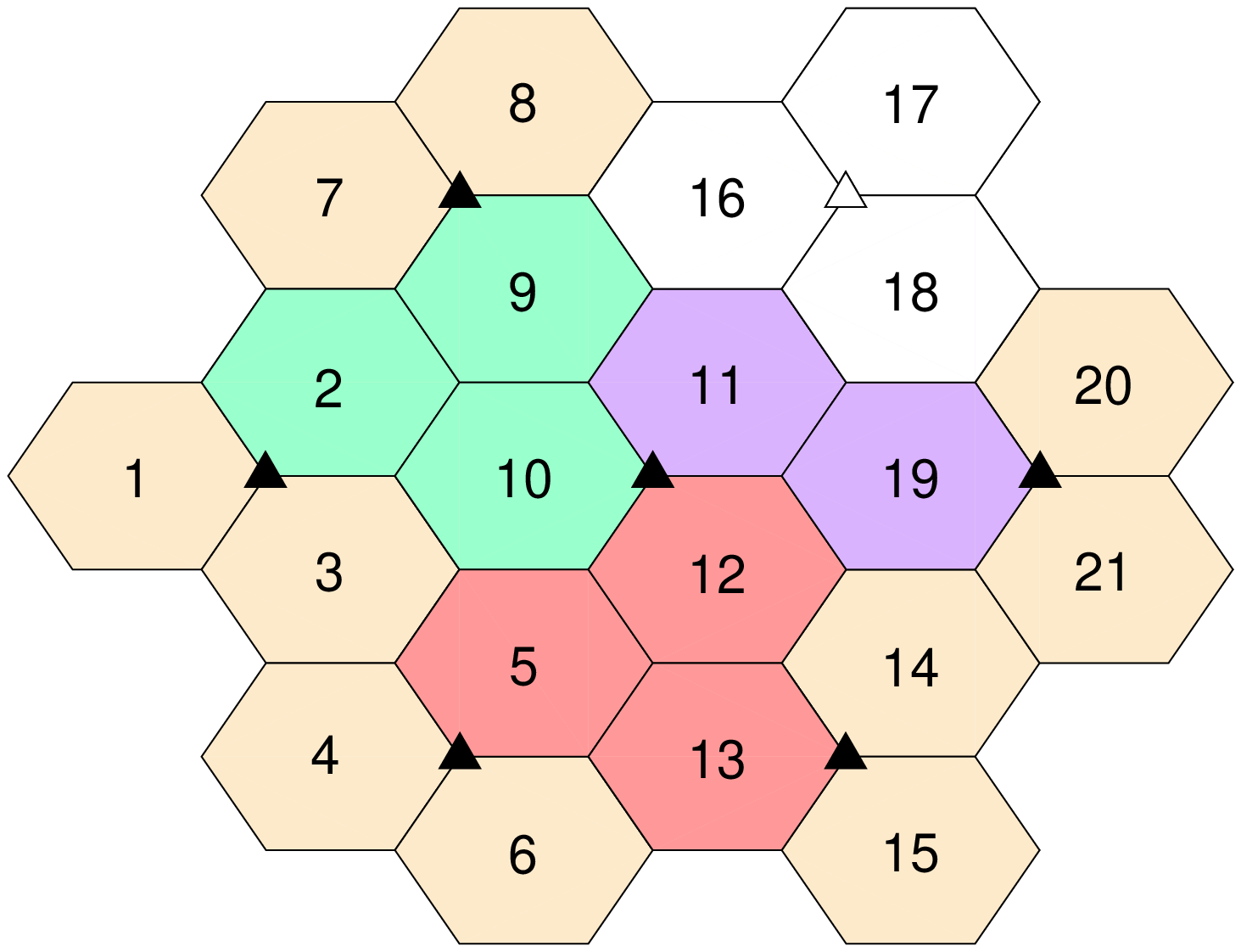}
\caption{$\mathcal{Z}_{1/7}$}
\label{fig:z2}
\end{subfigure}
\begin{subfigure}{0.35\textwidth}
\includegraphics[width=0.9\linewidth, height=5cm]{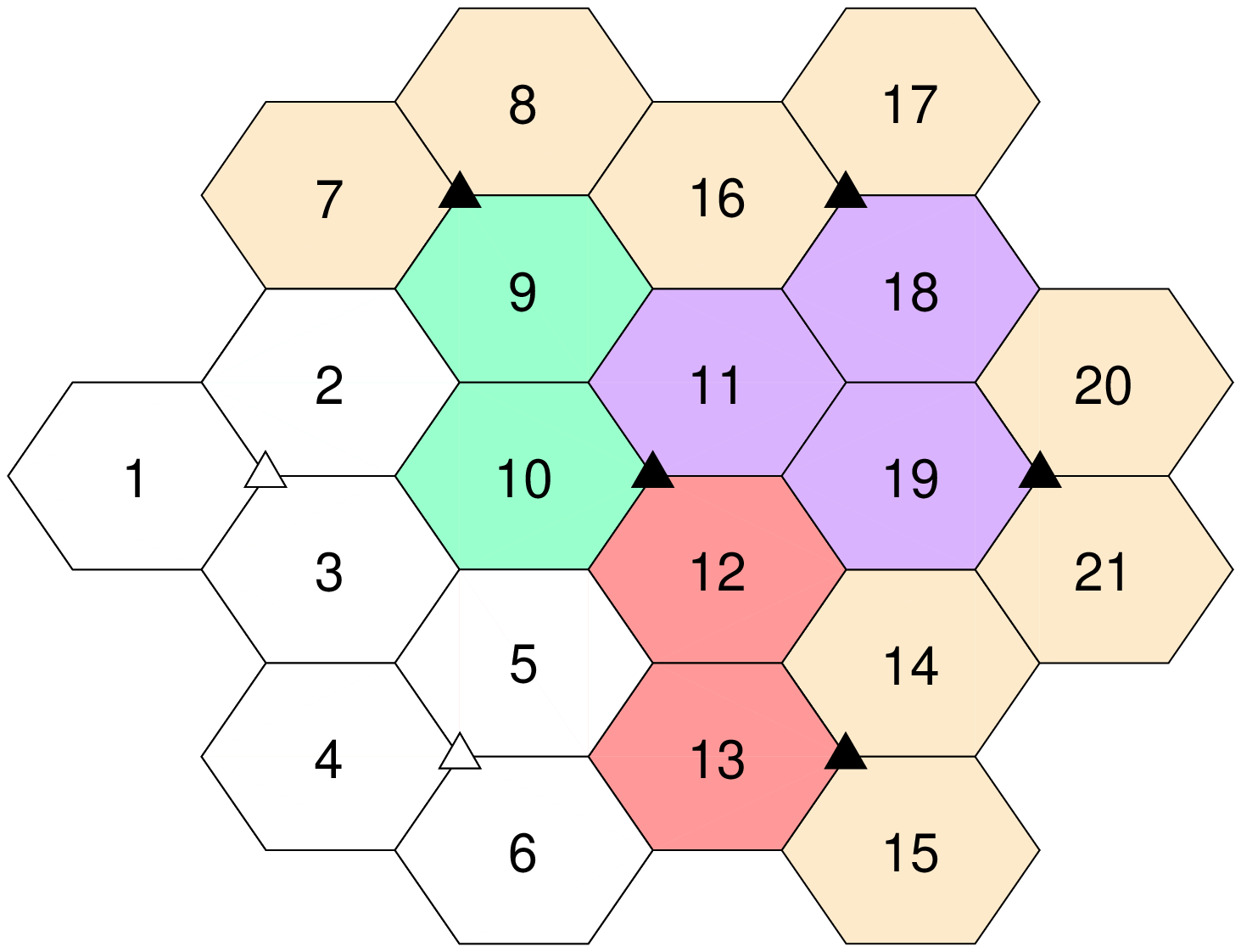}
\caption{$\mathcal{Z}_{2/7}$}
\label{fig:z3}
\end{subfigure}
\begin{subfigure}{0.35\textwidth}
\includegraphics[width=0.9\linewidth, height=5cm]{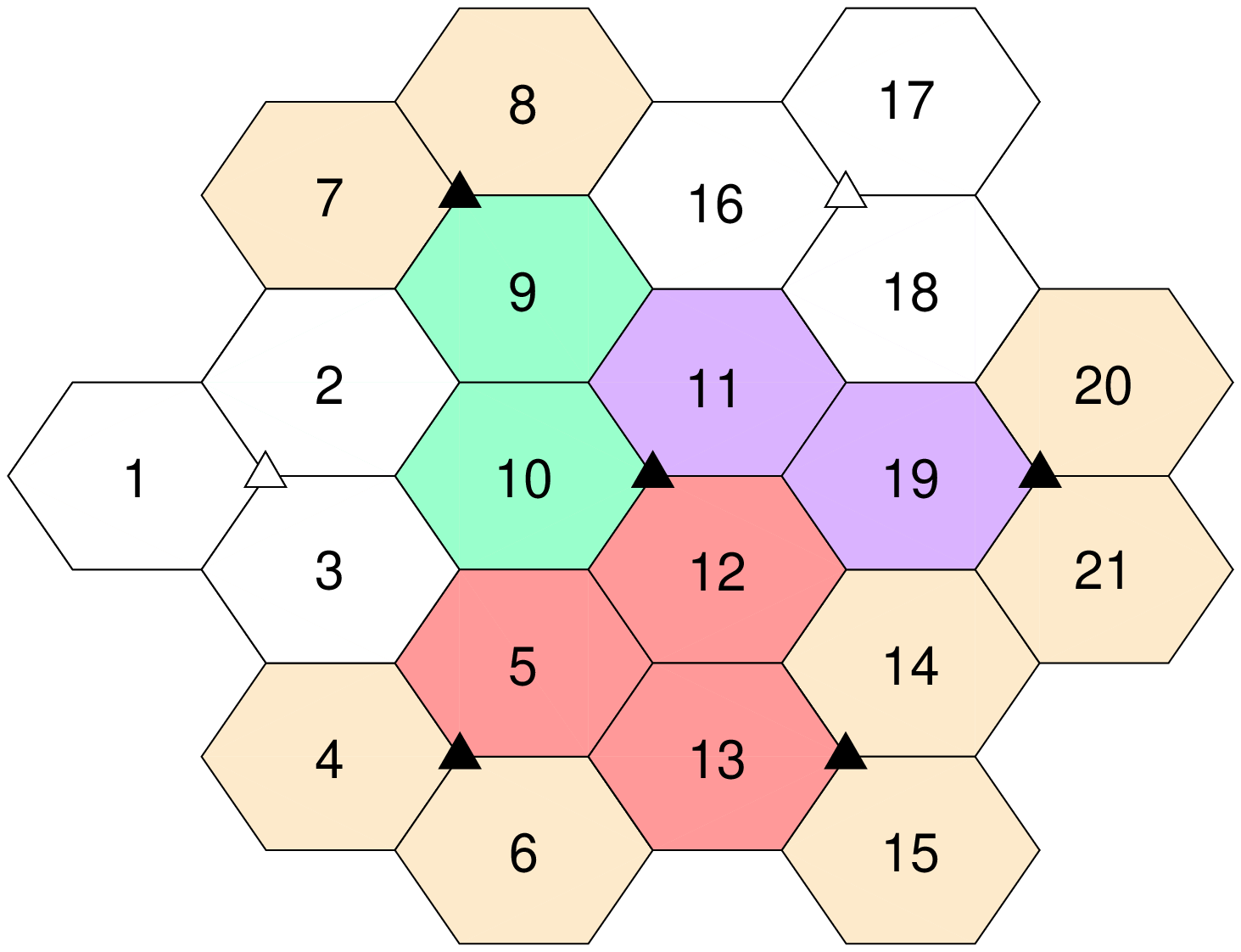} 
\caption{$\mathcal{Z}_{2/7}$}
\label{fig:z4}
\end{subfigure}
\begin{subfigure}{0.35\textwidth}
\includegraphics[width=0.9\linewidth, height=5cm]{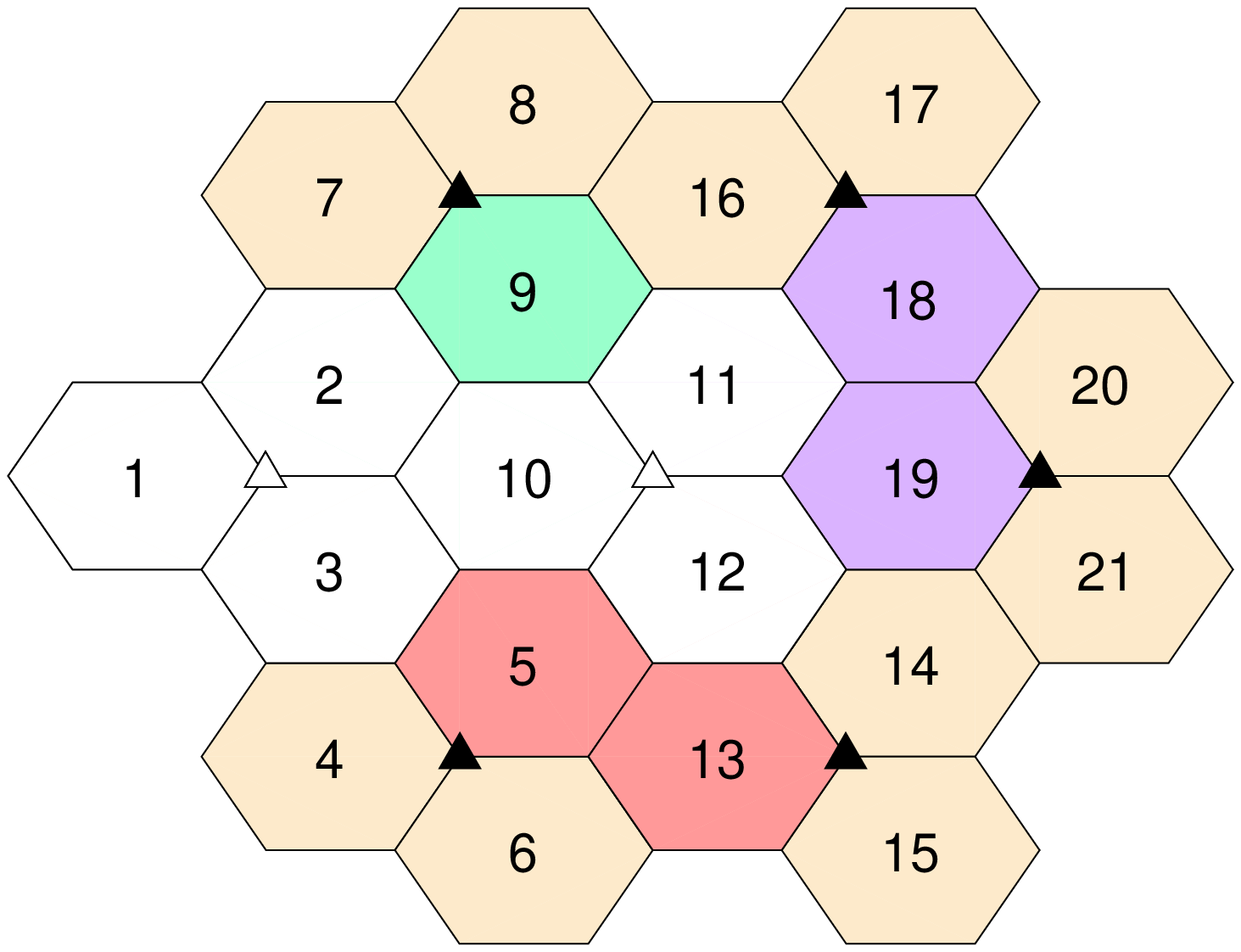}
\caption{$\mathcal{Z}_{2/7}$}
\label{fig:z5}
\end{subfigure}
\begin{subfigure}{0.35\textwidth}
\includegraphics[width=0.9\linewidth, height=5cm]{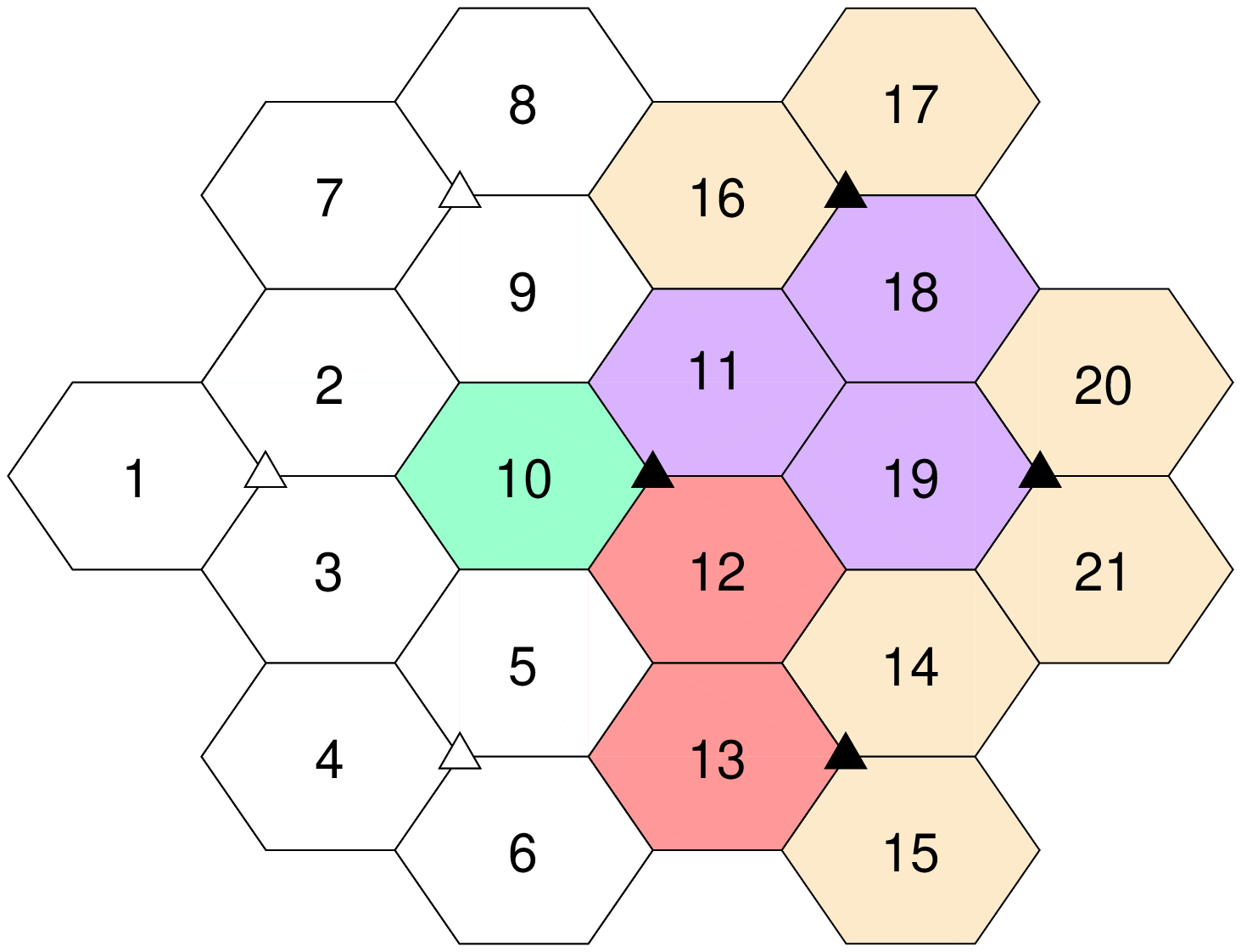}
\caption{$\mathcal{Z}_{3/7}$}
\label{fig:z6}
\end{subfigure}
\begin{subfigure}{0.35\textwidth}
\includegraphics[width=0.9\linewidth, height=5cm]{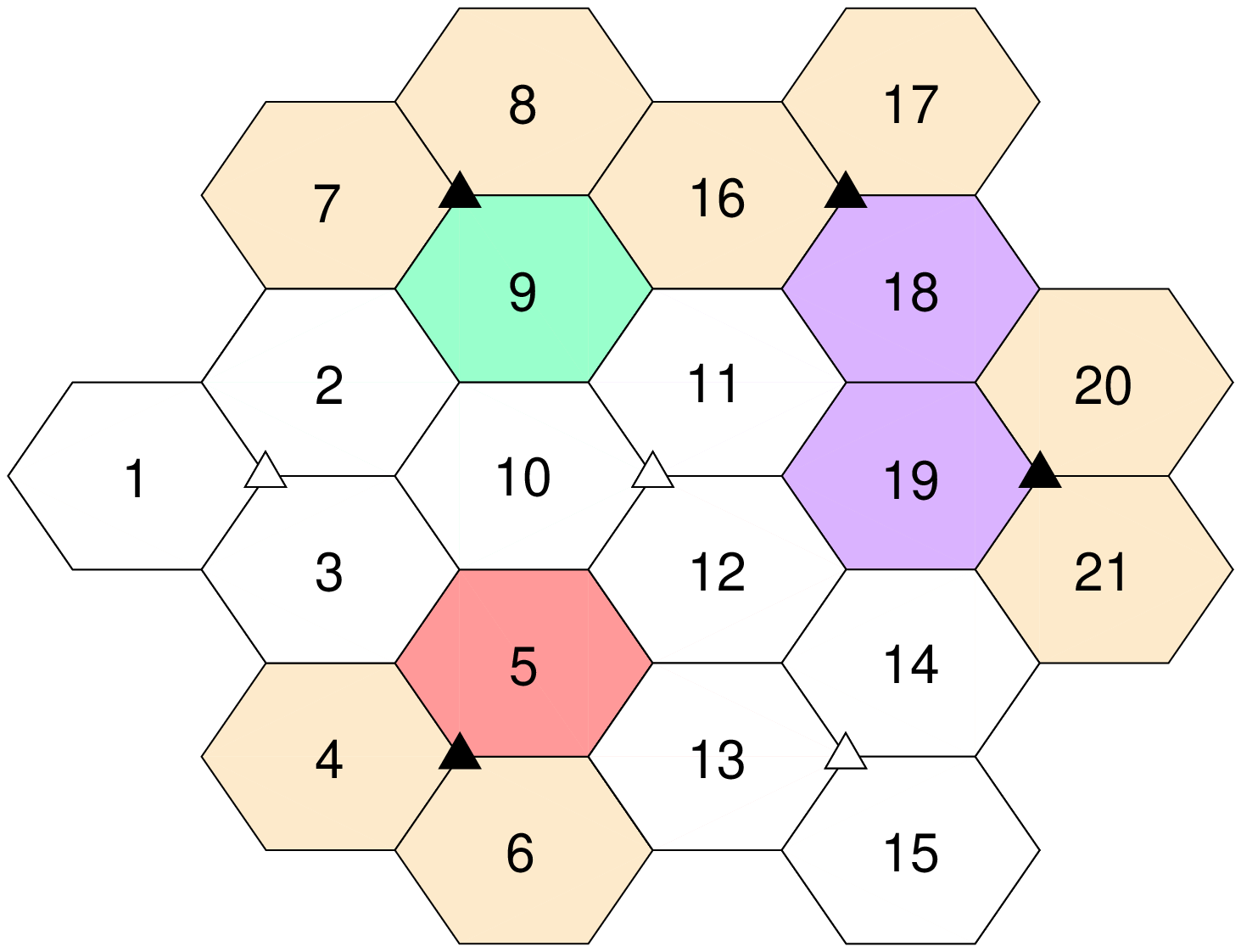}
\caption{$\mathcal{Z}_{3/7}$}
\label{fig:z7}
\end{subfigure}
\begin{subfigure}{0.35\textwidth}
\includegraphics[width=0.9\linewidth, height=5cm]{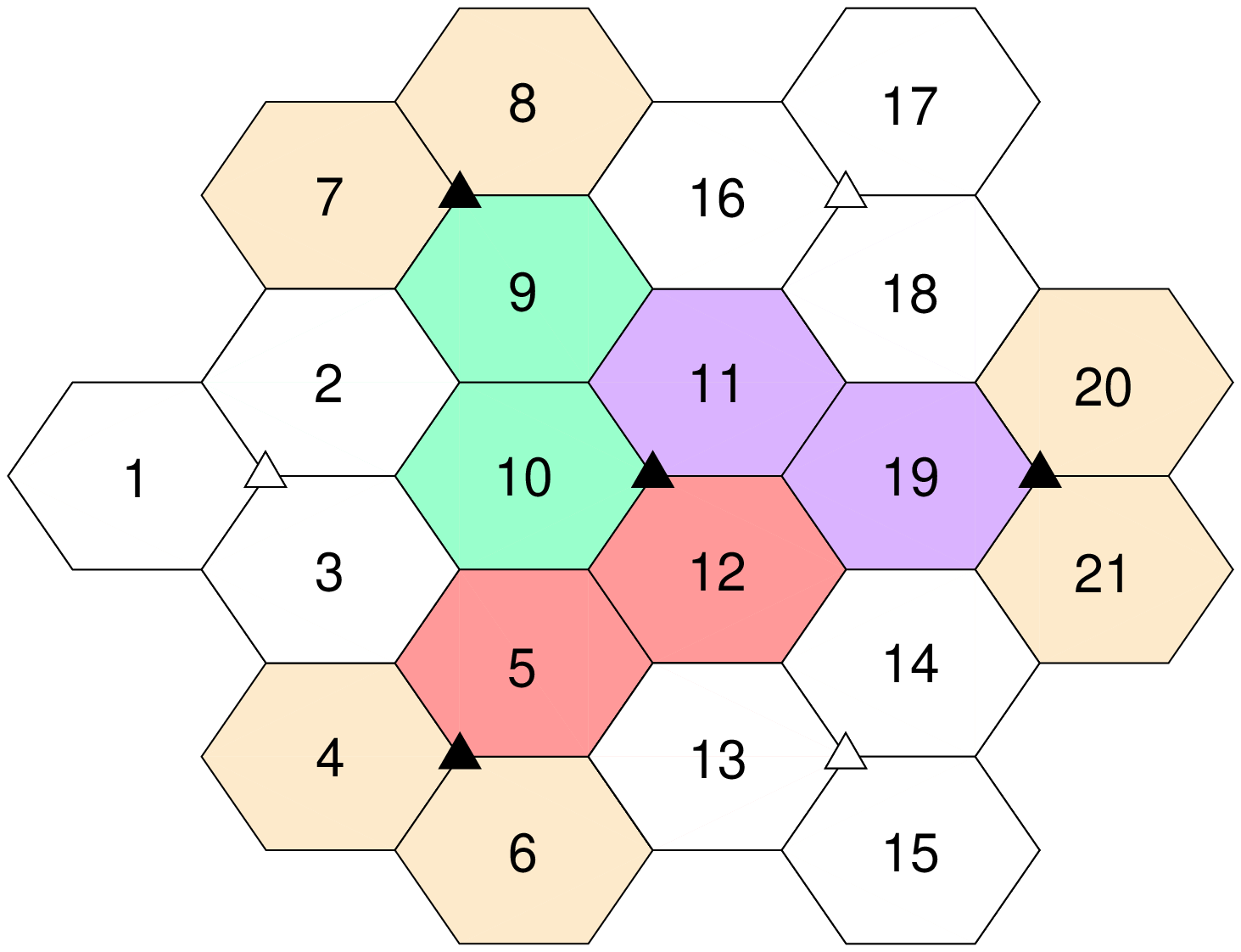}
\caption{$\mathcal{Z}_{3/7}$}
\label{fig:z8}
\end{subfigure}
\begin{subfigure}{0.35\textwidth}
\includegraphics[width=0.9\linewidth, height=5cm]{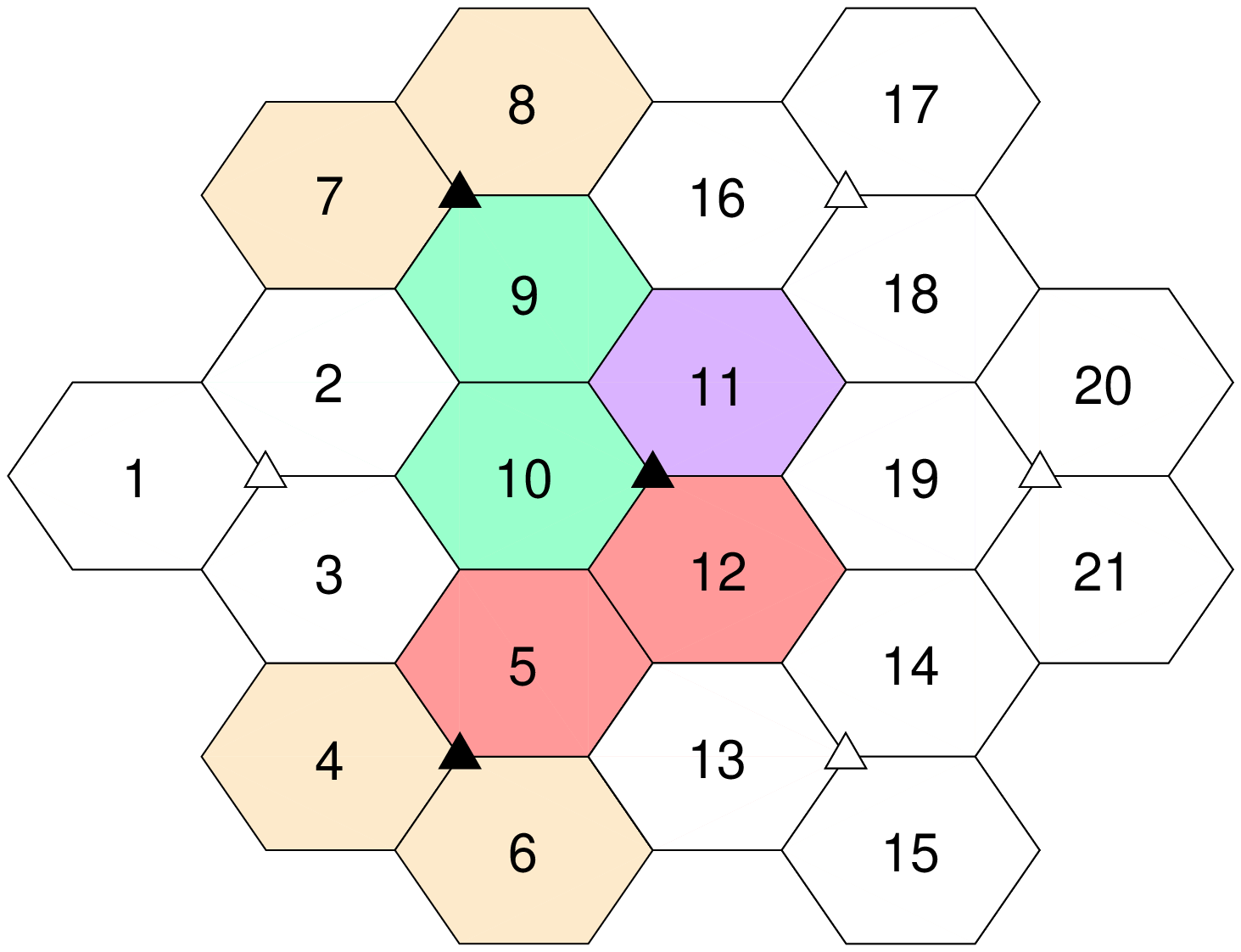}
\caption{$\mathcal{Z}_{4/7}$}
\label{fig:z9}
\end{subfigure}
\caption{Various BSS patterns for the center cluster for CoMP
configuration 3 (the solid black triangles represent BSs in ON state and white triangles represents BSs that are in OFF state).}
\label{fig:bss_patterns}
\end{figure*}

\subsection{CoMP}
We consider that
the sectors are grouped in pre-determined CoMP clusters
such that only sectors from the same CoMP cluster
can cooperate and perform CoMP.
This is a reasonable assumption
as CoMP requires a direct backhaul link
between participating sectors.
We denote the set of CoMP clusters by $\mathcal{Q}=\{1, 2, ...,Q\}$.
Without loss of generality, we focus on the center cluster
in Fig.~\ref{fig:1} represented by $q$
such that $\mathcal{B}_{q}$, $\mathcal{W}_{q}$,
and $\mathcal{V}_{q}$ denote the set of BSs, sectors, and
users in the cluster $q$, respectively.
Within the cluster $q$,
several configurations are possible for CoMP
based on which sectors perform CoMP together. We represent set of CoMP sectors present in a cluster $q$ as virtual clusters, which is represented by
$\mathcal{K}_{q}=\{1, 2, ...,K\}$.
In a virtual cluster $k$,
we use $\mathcal{S}_{k}$ and $\mathcal{U}_{k}$
to represent the set of sectors and users, respectively.
Thus, a cluster is a group of BSs that performs CoMP, and virtual cluster is the group of sectors within a cluster which performs CoMP.
Thus, $\mathcal{S}_{k} \subseteq \mathcal{W}_{q} \subset \mathcal{S} $.   
We consider the following
three possible CoMP
configurations in the cluster $q$.
\begin{itemize}
\item {{Configuration 1}}:
In this configuration, also referred to as $C_1$,
as shown in Fig.~\ref{fig:centre_c},
a CoMP user in cluster $q$ receive signals
jointly from a sectors $s$ of each BS in
the cluster $q$.
Thus, the virtual cluster is of size
$|\mathcal{W}_{q}|/3$ for $C_1$
\item {Configuration 2}:  In $C_2$, at most two sectors coordinate with each other as
shown in Fig.~\ref{fig:sch1}. Thus, 
sectors $1$, $15$, and $17$ do not perform CoMP,
while all the other sectors perform CoMP pairwise (sectors with the same colors cooperate).
\item {Configuration 3}: In Fig.~\ref{fig:sch2}, the Configuration 3 or $C_3$ is
presented. The sectors in sets of three
namely, $\{2,\ 9,\  10\}$, $\{5,\ 12,\ 13\}$, and $\{11\ , 18,\  19\}$
perform CoMP and the other
sectors in the cluster $q$ operate without CoMP
in $C_3$.
\end{itemize}
To focus on other aspects like user scheduling and resource
allocation for energy saving we have considered a
cluster of 7 BSs and only three CoMP configurations.
However, both the cluster size and the CoMP configurations
can be adapted for a practical system.
The sectors present in any virtual
cluster $\mathcal{S}_{k}$ will vary
based on the configuration under consideration
as shown in Fig.~\ref{fig:c_c}. 

We consider that the CoMP based system
allocates a fraction of time
for CoMP users in which the sectors in the
virtual cluster transmit jointly on
the downlink to the CoMP users.
Whenever the SINR of a user  $u$ associated to a sector $s$, in the virtual cluster $k$,
is less than a predetermined SINR threshold $\Gamma_d$, 
the user is served as a CoMP user.
Let $\theta_k$ denote the
time fraction in which such CoMP users
receive data jointly from their virtual cluster
$k$.
During the remaining downlink time
fraction $(1-\theta_k)$,
each sector transmits to the typical
non-CoMP users individually.
Note that each virtual cluster $k$ has
its own $\theta_k$.

In the CoMP time fraction $\theta_k$,
the downlink SINR received by a
user $u$ from any virtual cluster $k$ of
over subchannel $m$
(denoted by $\gamma_{u,k}^m$) is given by
\begin{equation}
	\gamma_{u,k}^m=\frac{ \sum\limits_{ {v \in \mathcal{S}_{k}} } P_{v}^m h_{u,v}^m}{\sum\limits_{ \substack { {\hat{v}} \in \mathcal{S} \\ \hat{v} \notin \mathcal{S}_{k} }} P_{\hat{v}}^mh_{u,\hat{v}}^m  + \sigma^2} \, ,
	\label{eq:SINR_k} 	
\end{equation}
where, $\sum\limits_{v \in \mathcal{S}_{k}} P_{v}^m h_{u,v}^m$ is the sum of the received powers
for user $u$
from all the sectors in the virtual cluster $k$
and $\sum\limits_{ \substack { {\hat{v}} \in \mathcal{S} \\ \hat{v} \notin \mathcal{S}_{k} }} P_{\hat{v}}^mh_{u,\hat{v}}^m$
is the interference from all the
other sectors in the system which are not part of this virtual cluster $k$.
Note that the SINR
for users associated with
the non-CoMP sectors and non-CoMP users of CoMP sectors
of cluster $q$ will be as in (\ref{eq: sinr_dl}).
The link rate for 
a CoMP user $u$ from a virtual cluster $k$
can be obtained using (\ref{eq:l_r}) and (\ref{eq:SINR_k}) as
\begin{equation}\label{eq:ruk}
	r_{u,k} =\frac{\eta(\gamma_{u,k}^m) \, SC_{OFDM} \, SY_{OFDM}}{T_{sc}}M \, .
\end{equation}
Next, we present the various BSS patterns considered in this work.

\subsection{BSS Patterns}
Let $\mathcal{Z}_{a1/a2}$ denote a BSS pattern
in which a1 out of the total a2 BSs in the cluster
are switched off. Hence, if a1 is equal to 0,
then all BSs in the cluster are active.
In Fig.~\ref{fig:bss_patterns},
we depict some of the possible
BSS patterns corresponding to
$\mathcal{Z}_{1/7}$, $\mathcal{Z}_{2/7}$, $\mathcal{Z}_{3/7}$, and $\mathcal{Z}_{4/7}$
for CoMP configuration $C_3$. The shaded black triangles represent active BSs and
white triangles represent
the BSs that have been switched off
in Fig.~\ref{fig:bss_patterns}.
We use idle and active states of
the BSs with OFF and ON state
interchangeably throughout the text. 
Note that Fig.~\ref{fig:c_c}c
represents $\mathcal{Z}_{0}$
for $C_3$,
where all BSs are active. For a given $a1$ in $\mathcal{Z}_{a1/a2}$,
multiple possible BSS patterns exist.
For example, Fig.~\ref{fig:z1} and Fig.~\ref{fig:z2} are both for
$\mathcal{Z}_{1/7}$. Seven such
combinations are possible for $\mathcal{Z}_{1/7}$
in which any one of the seven BS in the cluster can be switched off. 
The proposed optimization
problem and the solution heuristic
are valid for all such combinations.

%To avoid confusion and focus on
%the resource allocation we consider
%Fig.~\ref{fig:z2}, Fig.~\ref{fig:z4}, Fig.~\ref{fig:z8},
%and Fig.~\ref{fig:z9} as
%$\mathcal{Z}_{1/7}$, $\mathcal{Z}_{2/7}$, $\mathcal{Z}_{3/7}$, and $\mathcal{Z}_{4/7}$, respectively.
%
%Hence, we denote the mode of operation in a BSS pattern
%$\mathcal{Z}_{a1/a2}$ by $y$ such that $\mathcal{Z}_{a1/a2}^{y}$
%represents the various BSS patterns possible and $y \in \{1,2,..,^{a2}C_{a1}\}$. 
%Hence, we use $\mathcal{Z}_{a1/a2}$ and $\mathcal{Z}_{a1/a2}^{y}$
%interchangeably throughout the text.

\subsection{Performance Metrics}
The three key system performance metrics
of a cellular network are rate, coverage,
and energy.
We measure the system performance
of user rates through
the $\alpha-$Fair throughput
obtained over a cluster $q$ as follows \cite{r19}
\begin{align}\label{eq:tput_eqn}
\mathcal{T}_{\alpha}= & \ \Bigg(\frac{1}{|\mathcal{V}_{q}|}\sum\limits_{u \in \mathcal{V}_{q}}\lambda_{u}^{1-\alpha} \Bigg)^\frac{1}{1-\alpha}, &\alpha>0, \ \alpha \neq 1 \, , \nonumber \\
                     & \ \Big(\prod\limits_{u \in \mathcal{V}_{q}}\lambda_{u}\Big)^{\frac{1}{|\mathcal{V}_{q}|}}, &\alpha=1                                           
 \, ,\end{align}
 where, $\alpha$ is the fairness parameter, $\lambda_u$ is as defined in (\ref{lambda}), and
 $\mathcal{V}_{q}$ is the set of users associated with the cluster $q$.

We define
SINR coverage as the probability of a random user $u$
receiving SINR $\gamma_{u,s}^m$
greater than the minimum SINR threshold in Table~\ref{tab:1}
from at least one sector $s$.
Further,
we define rate coverage as the
probability of a random user $u$
receiving rate $\lambda_{u}$
greater than the rate threshold $R$.
This rate threshold is a system parameter
that can be controlled by the operator.

We consider the percentage of energy saved,
represented by $\mathcal{E}$, as the metric
for energy efficiency.
For a given BSS pattern $\mathcal{Z}_{a1/a2}$
which means $a1$ out of $a2$ BSs are switched off,
the percentage energy saving is
\begin{equation}\label{eq:e_e}
\mathcal{E}=\frac{a1}{a2} \times 100 \, .
\end{equation}
Next, we consider
a snapshot based approach
and consider a user realization
for a given user density $\mu$.
We formulate the joint BSS and CoMP
as an optimization problem
for this user realization.

\section{Joint BSS and CoMP Problem Formulation}
We use $w_{b}$ as a binary BSS variable to denote
BS $b$ in ON $(w_{b}=0)$ or OFF $(w_{b}=1)$ state.
We focus on the cluster $q$ in the center
as depicted in Fig.~\ref{fig:1}.
The power consumption of a BS $b$ in idle and active state is given by $P_{idle}$ and $P_{tot}$, respectively.
Then, for a given user realization,
to achieve energy efficiency,
we should optimize the following objective function \cite{sur}
\begin{equation}
 \min_{w_{b}}\sum\limits_{b \in \mathcal{B}_{q}} w_{b}P_{idle}^b + (1-w_{b})P_{tot}^b  \, . \label{eq:1}
\end{equation}
The objective function in (\ref{eq:1}) simplifies
to $\ \min_{w_{b}}\sum\limits_{b \in \mathcal{B}_{q}}w_{b}(P_{idle}^{b} - P_{tot}^{b})$.
Given $P_{idle}^{b}$ is always less than $P_{tot}^{b}$,
for a homogeneous cellular environment,
(\ref{eq:1}) is equivalent to
$ \ \max_{w_{b}}\sum
\limits_{b \in \mathcal{B}_{q}} w_{b}$.
Let $x_{u,s}$ denote an
association variable of user $u$ with sector $s$
such that $x_{u,s} \in \{0,1\}$.
Then, the BSS with CoMP problem can be framed
as an optimization problem for a
given user realization as follows.

We consider a maximum SINR based user association and its corresponding binary association variable as $x_{u,s}$. Based on this if any user $u$ is associated to a sector $s$, this variable $x_{u,s}$ is set to $1$, otherwise $x_{u,s}$ is set to $0$. We use $z_{u,s}$ as a binary variable
that denotes whether the user $u$
associated to sector $s$ will
receive CoMP transmission from the virtual cluster
$k$ (such that $s \in \mathcal{S}_k $ and $z_{u,s}$=1) or will receive conventional
downlink transmission from the sector $s$ ($z_{u,s}$=0).
We set the value of $z_{u,s}$ as 1 if the
$\gamma_{u,s}^m$ is less than the CoMP
SINR threshold $\Gamma_d$.
Given the number of CoMP and
non-CoMP users,
the virtual cluster $k$
has to decide the optimal
CoMP time fraction $\theta_k$.
We define $\beta_{u,k}$ as
the time fraction of $\theta_k$
for which an individual CoMP user $u$
receives joint downlink transmission
from the virtual cluster $k$.
Further,
any user in the center cluster $q$
should obtain a rate higher than
a pre-determined rate threshold $R$
with or without CoMP 
from corresponding virtual cluster $k$ or sector $s$,
respectively. 
Then, given the
utility function in (\ref{eq:u_alp}),
the joint BSS with CoMP resource allocation and
user scheduling problem for a 
cluster $q$ can be formulated
as the following optimization problem.
\begin{eqnarray}
\mathbb{B} & : & \ \max_{\substack{w_{b}, \, \Gamma_{d}, \, \beta_{u,k}, \\ \beta_{u,s}, \, \theta_{k}}}\sum\limits_{b \in \mathcal{B}_{q}} w_{b} \label{eq:12} \\
\mbox{s.t.} & & \sum\limits_{b \in \mathcal{B}_{q}}w_{b}  \leq  |\mathcal{B}_{q}|-1 \, , \label{eq:21} \\
w_{b} & \in & \{0,1\}, \ \forall b \in \mathcal{B}_{q} \, ,  \label{eq:31} \\
\lambda_{u} & = & \Big[ \sum\limits_{k \in \mathcal{K}_{q}} \sum\limits_{s \in \mathcal{S}_{k}}(1-\theta_{k})x_{u,s}(1-z_{u,s})\beta_{u,s}r_{u,s} \ + \  \nonumber \\ & &
\sum\limits_{k \in \mathcal{K}_{q}} \sum\limits_{s \in \mathcal{S}_{k}} \theta_{k}x_{u,s}z_{u,s}\beta_{u,k}r_{u,k} \Big] > R \  \ \forall u \in \mathcal{V}_{q} \, ,\label{eq:17} \\
\gamma_{u,s}^m&=&\frac{(1-w_{\lceil s/3 \rceil}) P_{s}^mh_{u,s}^m}{\sum\limits_{\substack {\hat{s} \neq s \\ \hat{s} \in \mathcal{S}}}(1-w_{\lceil \hat{s}/3 \rceil}) P_{\hat{s}}^mh_{u,\hat{s}}^m + \sigma^2 } \, ,
	\label{gmod1}\\
\gamma_{u,k}^m&=&\frac{ \sum\limits_{ {v \in \mathcal{S}_{k}} }(1-w_{\lceil v/3 \rceil}) P_{v}^m h_{u,v}^m}{\sum\limits_{ \substack { {\hat{v}} \in \mathcal{S} \\ \hat{v} \notin \mathcal{S}_{k} }} (1-w_{\lceil \hat{v}/3 \rceil}) P_{\hat{v}}^mh_{u,\hat{v}}^m  + \sigma^2} \,  , 
\label{gmod2}\\
x_{u,s}& =& \begin{cases}
		1, & \text{if $s = \arg \max_s\{\gamma_{u,s}^m$\}},\\
		0, & \text{otherwise},
	 \forall u \in \mathcal{V}_q, \forall s \in \mathcal{W}_q ,\end{cases} \label{eq:15} \\
z_{u,s}&=&\begin{cases}
		1, & \text{if $\gamma_{u,s}^m\leq \Gamma_{d}x_{u,s},  \ s \in \mathcal{S}_{k}, \ k \in \mathcal{K}_{q} \  s.t. \  |\mathcal{S}_{k}| > 1$}.\\
		0, & \text{otherwise}, \ \forall u \in \mathcal{V}_q, \forall s \in \mathcal{W}_q,	\end{cases}  \label{eq: bin_comp1} \\
&  \displaystyle\sum\limits_{s \in \mathcal{S}_{k}} & \sum_{u \in \mathcal{U}_{k}} z_{u,s}x_{u,s}\beta_{u,k}  \leq   1, \, \forall k \in \mathcal{K}_{q} \, ,\label{eq: time_c} \\
&   \displaystyle\sum\limits_{u \in \mathcal{U}_{k}} & (1-z_{u,s})x_{u,s}\beta_{u,s}   \leq    1 , \ \forall s \in \mathcal{S}_{k}, \, \forall k \in \mathcal{K}_{q}  \, , \ \label{eq: time_nc} \\
\beta_{u,s} & \geq & 0, \, \forall u \in \mathcal{U}_{k}, \ \forall s \in \mathcal{S}_{k}, \, \forall k \in \mathcal{K}_{q} \, , \label{eq:beta_z1} \\
\beta_{u,k} & \geq & 0, \  \forall u \in \mathcal{U}_{k}, \, \forall k \in \mathcal{K}_{q}  \, , \label{eq:beta_z2} \\
\theta_k &\in & [0,1], \, \forall k \in \mathcal{K}_{q} \, , 
	\label{eq: theta_range} \\
	\Gamma_d &\in & [\xi_{min}^d,\xi_{max}^d] \, ,
	\label{eq: th_range} 
\end{eqnarray}
where,
the objective function in (\ref{eq:12})
ensures maximum energy savings,
while, the constraint in (\ref{eq:21})
is to ensure that
atleast one BS in the center cluster is in ON state,
the constraint in (\ref{eq:31})
reflects that
a BS can be either in ON or OFF state,
the constraint in (\ref{eq:17}) is the resultant rate of a user with joint BSS and CoMP, 
the constraint in (\ref{gmod1}) is required
to account for the change in SINR from a sector
with BSS,
the SINR from virtual cluster $k$
is recomputed in the constraint in
(\ref{gmod2}) as
with BSS the received power from a
sector $v$ corresponding to BS $b=\lceil v/3 \rceil$
or received power
from an interfering sector $\hat{v}$ can be zero
if the corresponding BS is switched off,
the constraint in
(\ref{eq:15}) is required to re-compute
user association with BSS through
the additional term of $(1-w_b)$ that ensures the maximum
SINR is computed only over the BSs that are still in
ON state, the constraint in (\ref{eq: bin_comp1})
ensures that a user is served as a CoMP
user based on received SINR
only from sectors of BSs still in ON state
and for virtual cluster with
more than one sector available for CoMP,
the constraint in (\ref{eq: time_c}) indicates that time fractions
of $\theta_k$ allocated to all CoMP users in
cluster $k$ must be less than equal to 1.
Similarly, the constraint in
(\ref{eq: time_nc})
indicates that time fractions of $(1-\theta_k)$
allocated individually in each sector $s$
to non-CoMP users must be less than equal to 1.
The constraints in (\ref{eq:beta_z1}) and (\ref{eq:beta_z2})
are required to ensure non-negative time fractions for non-CoMP and CoMP users, respectively.
The constraint in (\ref{eq: theta_range})
ensures that the
CoMP time fraction is not more than the
total available time.
The values of $\xi_{min}^d$ and $\xi_{max}^d$ in the constraint (\ref{eq: th_range})
define the permitted range for the CoMP threshold $\Gamma_d$.

Note that the optimization problem
presented in (\ref{eq:12})
is an MINLP and the problem
becomes more complex with increasing number of BSs,
i.e., $|\mathcal{B}_q|$.
Therefore, we decompose
the joint problem of BSS and CoMP
in (\ref{eq:12})
into purely a CoMP
resource allocation and user scheduling problem
in the next section, and
use it to re-frame a simplified
BSS with CoMP problem later.

\section{CoMP Problem Formulation}
For the CoMP based system,
we use $z_{u,s}$, and $x_{u,s}$ as binary variables
as explained in the previous section.
For a given user realization, this CoMP problem jointly determines the solution for the optimal resource fraction $\theta_{k}$ that can be allocated for CoMP users, the optimal time fraction scheduled for individual users, i.e., $\beta_{u,s}(1-\theta_{k})$ fraction of time that can be allocated to a non-CoMP user $u$ by a sector $s$, and $\beta_{u,k}\theta_{k}$ fraction of time that can be allocated to a CoMP user jointly from the sectors in the virtual cluster $k$.
%that denotes whether the user $u$
%associated to sector $s$ will
%receive CoMP transmission from the virtual cluster
%$k$ (such that $s \in \mathcal{S}_k $ and $z_{u,s}$=1) or will receive conventional
%downlink transmission from the sector $s$ ($z_{u,s}$=0).
%We set the value of $z_{u,s}$ as 1 if the
%$\gamma_{u,s}^m$ is less than the CoMP
%SINR threshold $\Gamma_d$.
%Given the number of CoMP and
%non-CoMP users,
%the virtual cluster $k$
%has to decide the optimal
%CoMP time fraction $\theta_k$.
%Further, we define $\beta_{u,k}$ as
%the time fraction of $\theta_k$
%for which an individual CoMP user $u$
%receives joint downlink transmission
%from the virtual cluster $k$.
Then, given the
utility function in (\ref{eq:u_alp}),
the joint CoMP resource allocation and
user scheduling problem for a virtual
cluster $k$ can be formulated
as the following optimization problem.
\begin{eqnarray}
\mathbb{P} \ & : &  \ \ \max_{\substack{\Gamma_{d}, \, \theta_k,\\
\beta_{u,s}, \, \beta_{u,k}}} \sum\limits_{u \in \mathcal{U}_{k}} \mathbb{U}_{\alpha}(\lambda_u) \, , \label{eq: problem_dl} \\
\mbox{s.t. } \lambda_{u} & = & (1-\theta_{k}) \sum\limits_{s \in S_{k}}x_{u,s}(1-z_{u,s})\beta_{u,s}r_{u,s} \ + \  \nonumber \\ & &
\theta_{k}\sum\limits_{s \in S_{k}} x_{u,s}z_{u,s}\beta_{u,k}r_{u,k},  \  \ \forall u \in \mathcal{U}_{k} \, ,\label{eq: user_rate}
\end{eqnarray}
\begin{eqnarray}
x_{u,s}& =& \begin{cases}
		1, & \text{if $s = \arg \max_s\{\gamma_{u,s}^m$\}},\\
		0, & \text{otherwise},
	 \forall u \in \mathcal{U}_k, \forall s \in \mathcal{S}_k \, ,\end{cases} \label{eq:001}\\
z_{u,s} & = &\begin{cases}
		1, & \text{if $\gamma_{u,b}^m\leq \Gamma_{d}x_{u,s}, \ s \in \mathcal{S}_{k} \  s.t. \  |\mathcal{S}_{k}| > 1$},\\
		0, & \text{otherwise}, \ \forall u \in \mathcal{U}_k, \forall s \in \mathcal{S}_k \, ,
	\end{cases}         \label{eq: bin_comp}	 \\ 
&  \displaystyle\sum\limits_{s \in \mathcal{S}_{k}} & \sum_{u \in \mathcal{U}_{k}} z_{u,s}x_{u,s}\beta_{u,k}  \leq   1 \, ,\label{eq: c_time_c} \\
&   \displaystyle\sum\limits_{u \in \mathcal{U}_{k}} & (1-z_{u,s})x_{u,s}\beta_{u,s}   \leq    1 , \ \forall s \in \mathcal{S}_{k}  \, , \ \label{eq: c_time_nc} \\
\beta_{u,s} & \geq & 0, \, \forall u \in \mathcal{U}_{k}, \ \forall s \in \mathcal{S}_{k} \, , \label{eq:c_beta_z1} \\
\beta_{u,k} & \geq & 0, \  \forall u \in \mathcal{U}_{k} \, , \label{eq:c_beta_z2} \\
\theta_k &\in & [0,1] \, , 
	\label{eq: c_theta_range} \\
& & (\ref{eq: sinr_dl}) \, ,(\ref{eq: th_range}) \, , \nonumber		
\end{eqnarray}
where, the user rate is defined by (\ref{eq: user_rate})
such that any non-CoMP users $u$ gets a fraction of $\beta_{u,s}(1-\theta_k)$ from the sector $s$
and any CoMP users $u$ gets a fraction of $\beta_{u,k} \theta_k$ from
all sectors in $k$,
$x_{u,s}$ in (\ref{eq:001})
represents the maximum SINR based binary user association
variable,
the constraint in (\ref{eq: bin_comp}) implies
that a user can be either CoMP or non-CoMP with corresponding
binary $z_{u,s}$. 
The $r_{u,k}$ in (\ref{eq: user_rate})
is given in (\ref{eq:ruk}).

The joint resource allocation and user
scheduling problem in (\ref{eq: problem_dl})
is also an MINLP
which is difficult to solve simultaneously
for the multiple optimization
variables (namely, $\Gamma_{d}$, $\theta_k$, $\beta_{u,s}$, $\beta_{u,k}$).
Hence, we next present propositions 
that provide optimal solutions
with respect to $\beta_{u,s}$, $\beta_{u,k}$,
and $\theta_k$ for a given $\Gamma_{d}$ and $x_{u,s}$
in a virtual cluster $k$.
This is a valid assumption
as user association ($x_{u,s}$)
is typically maximum SINR based
and thresholds like $\Gamma_{d}$
can be determined via simulations.
We first present Proposition 1 which solves the user scheduling problem for any CoMP resource allocation ($\theta_{k}$) because the user scheduling is independent $\theta_{k}$. %Further, the next Proposition 2 solves for  optimal CoMP resource allocation ($\theta_{k}$ given the optimal CoMP user schedulers $\beta_{u,s}$, and $\beta_{u,k}$.
\begin{prop}
For a virtual cluster $k$,
given a user association $x_{u,s}$,
a CoMP SINR threshold $\Gamma_{d}$,
at least one CoMP user with $\gamma_{u,s}^m \leq\Gamma_{d}$,
and any CoMP time fraction $\theta_k$,
the optimal time fraction of $(1-\theta_k)$,
allocated by the $\alpha$-Fair scheduler in
any sector $s\in\mathcal{S}_k$ for a non-CoMP user $u$ is equal to
\begin{equation}\label{eq:aa1}
\beta_{u,s}^* = \dfrac{\mathtt{t}_{u,s,\alpha}}{\sum\limits_{v \in \mathcal{U}_{nc,s}}\mathtt{t}_{v,s,\alpha}} \ , \forall	s \in \mathcal{S}_{k}, \ \forall u \in \mathcal{U}_{nc} \, ,
\end{equation}
where, $\mathtt{t}_{u,s,\alpha}=r_{u,s}^{\frac{1-\alpha}{\alpha}}$,
and the optimal time fraction of $\theta_k$ allocated
by an $\alpha$-Fair scheduler for
all the sectors jointly to a CoMP user $u$ is equal to
\begin{equation}\label{eq:aa2}
\beta_{u,k}^* = \dfrac{\mathtt{t}_{u,k,\alpha}}{\sum\limits_{v \in \mathcal{U}_{c}}\mathtt{t}_{v,k,\alpha}}, \ \forall u \in \mathcal{U}_{c} \, ,
\end{equation}
where, $\mathtt{t}_{u,k,\alpha}=r_{u,k}^{\frac{1-\alpha}{\alpha}}$,
$\mathcal{U}_{c}=\{1,2, ...U_{c}\}$,
$\mathcal{U}_{nc}=\{1,2, ...U_{nc}\}$,
and $\mathcal{U}_{nc,s}=\{1,2, ...U_{nc,s}\}$
denote the set of CoMP users in $\mathcal{S}_k$,
the set of non-CoMP users in $\mathcal{S}_k$,
and the set of non-CoMP users in any sector $s\in\mathcal{S}_k$
in the virtual cluster, respectively.
\end{prop}
\emph{Proof:}
For any given user association $x_{u,s}$
(note that it need not be maximum SINR based)
and CoMP SINR threshold $\Gamma_{d}$,
the virtual cluster $k$ can compute $z_{u,s}$ using (\ref{eq: bin_comp}).
Given binary $z_{u,s}$, a user $u$
can be classified as CoMP or non-CoMP user into the sets 
$\mathcal{U}_{c}$ or $\mathcal{U}_{nc}$, respectively.
Further, the set of non-CoMP users
for every sector $s\in\mathcal{S}_k$, denoted by
$\mathcal{U}_{nc,s}$, can be obtained.
Then, as $\mathcal{U}_{k}=\mathcal{U}_{c}\cup \mathcal{U}_{nc}$,
the objective function in (\ref{eq: problem_dl})
denoted by $\mathbb{Y}$ can be represented as
\begin{equation}
\mathbb{Y}=\sum_{u \in \mathcal{U}_{k}}\dfrac{\lambda_u^{1-\alpha}}{1-\alpha} = \sum_{u \in \mathcal{U}_{nc}}\dfrac{\lambda_u^{1-\alpha}}{1-\alpha} + \sum_{u \in \mathcal{U}_{c}}\dfrac{\lambda_u^{1-\alpha}}
{1-\alpha} \, ,
\end{equation}
which using (\ref{eq: user_rate}) becomes
\begin{eqnarray}
\mathbb{Y}&=&\sum\limits_{u \in \mathcal{U}_{nc}}
\sum \limits_{s \in S_{k}} x_{u,s} \dfrac{(1-\theta_{k})^{1-\alpha}(r_{u,s}\beta_{u,s})^{1-\alpha}}{1-\alpha} \nonumber \\
&+& \sum\limits_{u \in \mathcal{U}_{c}}\sum \limits_{s \in S_{k}} x_{u,s} \dfrac{\theta_{k}^{1-\alpha}(r_{u,k}\beta_{u,k})^{1-\alpha}}{1-\alpha} \, .
\nonumber
\end{eqnarray}
Then,
for any given $\theta_k$, $x_{u,s}$, and $\Gamma_{d}$,
the optimization problem in (\ref{eq: problem_dl})
can be simplified to
\begin{eqnarray}
\mathbb{P}^* \ &:&  \ \ \max_{\substack{\beta_{u,s},\beta_{u,k}}} \ \ \mathbb{Y} \label{eq: mp1} \\
\mbox{s.t.}& &\sum\limits_{u \in \mathcal{U}_{nc,s}} \beta_{u,s} \leq 1, \ \forall s \in \mathcal{S}_{k}  \, ,
\label{eq: mp2} \\
& & \sum\limits_{u \in \mathcal{U}_{c}} \beta_{u,k} \leq 1 \, ,
\label{eq: mp3}\\
& &
(\ref{eq:beta_z1}), \mbox{ and } (\ref{eq:beta_z2}) \, , \nonumber 
\end{eqnarray}
where, (\ref{eq: mp2}) and (\ref{eq: mp3}) are obtained
from (\ref{eq: time_c}) and (\ref{eq: time_nc}), respectively.
The Lagrangian function of (\ref{eq: mp1})
can be defined as 
\begin{eqnarray}
\mathbf{L}(\mathbb{Y},\mathbb{V}_s, \mathbb{V}_{k},\mathbb{X}_{u,s},\mathbb{X}_{u,k}) = -\mathbb{Y}+\sum\limits_{s \in \mathcal{S}_{k}}\mathbb{V}_{s}\Big(\sum\limits_{u \in \mathcal{U}_{nc,s}}\beta_{u,s}-1\Big) + \nonumber \\
 \mathbb{V}_{k}\Big(\sum\limits_{u \in \mathcal{U}_{c}}\beta_{u,k}-1\Big) 
- \sum\limits_{s \in \mathcal{S}_{k}} \sum\limits_{u \in \mathcal{U}_{nc,s}}\mathbb{X}_{u,s}\beta_{u,s} - \sum\limits_{u \in \mathcal{U}_{c}}\mathbb{X}_{u,k}\beta_{u,k} \, ,\label{eq: mp5}
\end{eqnarray}
where, $\mathbb{V}_s$, $\mathbb{V}_{k}$, $\mathbb{X}_{u,s}$,
and $\mathbb{X}_{u,k}$ are the KKT multipliers \cite{KKT}
for (\ref{eq: mp2}), (\ref{eq: mp3}), (\ref{eq:beta_z1}),
and (\ref{eq:beta_z2}), respectively.
Considering the complementary slackness KKT conditions,
the values of $\mathbb{X}_{u,s}$ and $\mathbb{X}_{u,k}$
turn out to be zero
for a user $u$ whenever it receives non-zero
$\beta_{u,s}$ or $\beta_{u,k}$ from
a sector $s$ or a cluster $k$, respectively.
Thus, the corresponding (\ref{eq: mp5})
for users receiving non-zero rate
(i.e., $x_{u,s}$=1)
becomes
\begin{equation}\label{eq: mp6}
\mathbf{L}(\mathbb{Y},\mathbb{V}_s,\mathbb{V}_{k})= -\mathbb{Y}+\sum\limits_{s \in \mathcal{S}_{k}}\mathbb{V}_{s}\Big(\sum\limits_{u \in \mathcal{U}_{nc,s}}\beta_{u,s} \ - \ 1 \Big) + \mathbb{V}_{k}\Big(\sum\limits_{u \in \mathcal{U}_{c}}\beta_{u,k} \ - \ 1 \Big) \, .
\end{equation}
The first-order stationarity
conditions of (\ref{eq: mp6})
for (\ref{eq: mp2}) and (\ref{eq: mp3}) result in
\begin{eqnarray}
\dfrac{d\mathbf{L}}{d\beta_{u,s}}& =& - \left[(1-\theta_{k})r_{u,s}\right]^{1-\alpha}\beta_{u,s}^{-\alpha} + \mathbb{V}_{s} =0\label{eq: mp7}  \mbox{ and}\\
\dfrac{d\mathbf{L}}{d\beta_{u,k}}& =& - \left[\theta_{k} r_{u,k}\right]^{1-\alpha}\beta_{u,k}^{-\alpha} + \mathbb{V}_{k} =0
\, , \mbox{ respectively.}
\label{eq: mp8}
\end{eqnarray}
Solving (\ref{eq: mp7}) and (\ref{eq: mp8})
jointly with (\ref{eq: mp2}) and (\ref{eq: mp3})
result in (\ref{eq:aa1}) and (\ref{eq:aa2}),
respectively.
This completes the proof of $Proposition$ 1. \hfill \rule{1.5ex}{1.5ex}

Note that for $\alpha=1$, i.e., a proportional fair scheduler,
(\ref{eq:aa1}) and (\ref{eq:aa2})
result in time fractions ${1}/{N_{nc,s}}$
and ${1}/{N_c}$ for Non-CoMP and CoMP users,
respectively, in any sector $s$ of the CoMP cluster. The result presented in (\ref{eq:aa1}) and (\ref{eq:aa2}) gives the time fraction
allocated to the set of users $\mathcal{U}_{k}$ in the cluster $k$. It is observed from (\ref{eq:aa1}) that the time fraction
allocated for a non-CoMP user $u$ depends only
on the non-CoMP users in the sector $s$.
Further, (\ref{eq:aa2}) presents
the time fraction allocated for a CoMP user $u$ in the virtual cluster $k$ which depends on all
CoMP users in the same virtual cluster $k$.
Next in Proposition 2, we present optimal resource allocation of CoMP users for the $\alpha$-Fair scheduler.

\begin{table}
	\begin{center}	
	\caption{Various values of $\alpha$ and corresponding $\theta_{k}^*$ for a virtual cluster $k$. }
		\label{tab:2}				
		\begin{tabular}{|c|c|c|} 
			\hline			
			$\alpha$ & $\delta$ & $\theta_{k}^*$ \\
			\hline			
			$1$ & $\dfrac{N_{c}}{N_{nc}}$ & $\dfrac{N_c}{N_{c}+N_{nc}}$ \\
			\hline
			$2$ & $\sqrt{\Bigg[\dfrac{\sum\limits_{u \in \mathcal{U}_{c}}(r_{u,k}\beta_{u,k}^*)^{-1}}{\sum\limits_{u \in \mathcal{U}_{nc}}\sum \limits_{s \in S_{k}} x_{u,s}(r_{u,s}\beta_{u,s}^*)^{-1}}\Bigg]}$ & $\dfrac{\delta}{1+\delta}$ \\
			\hline	
			$\alpha$ & $\sqrt{\Bigg[\dfrac{\sum\limits_{u \in \mathcal{U}_{c}}(r_{u,k}\beta_{u,k}^*)^{1-\alpha}}{\sum\limits_{u \in \mathcal{U}_{nc}}\sum \limits_{s \in S_{k}} x_{u,s}(r_{u,s}\beta_{u,s}^*)^{1-\alpha}}\Bigg]}$ & $\dfrac{\delta}{1+\delta}$\\
			\hline				
			\end{tabular}		
	\end{center}
\end{table}

\begin{prop}
For a given user association $x_{u,s}$
and CoMP SINR threshold $\Gamma_{d}$,
the optimal time fraction $\theta_{k}^{*}$
for CoMP users in a
virtual cluster $k$ is given by
\begin{equation}
\theta_{k}^{*}=\dfrac{\delta}{1+{\delta}} \, ,
\label{eq:astep3}
\end{equation}
where,
\begin{equation}
\delta=\left[\dfrac{\sum\limits_{u \in \mathcal{U}_{c}}(r_{u,k}\beta_{u,k}^*)^{1-\alpha}}{\sum\limits_{u \in \mathcal{U}_{nc}}\sum \limits_{s \in S_{k}} x_{u,s} (r_{u,s}\beta_{u,s}^*)^{1-\alpha}}\right]^{\dfrac{1}{\alpha}} \, , 
\end{equation}
with $\beta_{u,s}^*$ and $\beta_{u,k}^*$
as in (\ref{eq:aa1}) and (\ref{eq:aa2}), respectively.
\end{prop}
\emph{Proof:}
For any given user association $x_{u,s}$
and CoMP SINR threshold $\Gamma_{d}$,
the virtual cluster $k$ can be classify
users into the sets 
$\mathcal{U}_{c}$ or $\mathcal{U}_{nc}$
as shown in the proof of Proposition 1.
Then, as $\mathcal{U}_{k}=\mathcal{U}_{c}\cup \mathcal{U}_{nc}$,
the objective function in (\ref{eq: problem_dl})
can be represented as
\begin{equation}
\sum_{u \in \mathcal{U}_{k}}\dfrac{\lambda_u^{1-\alpha}}{1-\alpha} = \sum_{u \in \mathcal{U}_{nc}}\dfrac{\lambda_u^{1-\alpha}}{1-\alpha} + \sum_{u \in \mathcal{U}_{c}}\dfrac{\lambda_u^{1-\alpha}}
{1-\alpha} \, , \nonumber
\end{equation}
which given $x_{u,s}$ is binary,
(\ref{eq: user_rate}), (\ref{eq:aa1}), and
(\ref{eq:aa2}) becomes
\begin{equation}
\sum\limits_{u \in \mathcal{U}_{nc}}\sum \limits_{s \in S_{k}} x_{u,s}\dfrac{(1-\theta_{k})^{1-\alpha}(r_{u,s}\beta_{u,s}^*)^{1-\alpha}}{1-\alpha} + \sum\limits_{u \in \mathcal{U}_{c}}\dfrac{\theta_{k}^{1-\alpha}(r_{u,k}\beta_{u,k}^*)^{1-\alpha}}{1-\alpha} \, .
\label{eq:astep1}
\end{equation}
Differentiating (\ref{eq:astep1}) with respect to $\theta_k$
and equating to 0 gives
\begin{equation}
(1-\theta_{k}^*)^{-\alpha}\sum\limits_{u \in \mathcal{U}_{nc}}\sum \limits_{s \in S_{k}} x_{u,s}(r_{u,s}\beta_{u,s}^*)^{1-\alpha} = (\theta_{k}^*)^{-\alpha}\sum\limits_{u \in \mathcal{U}_{c}}(r_{u,k}\beta_{u,k}^*)^{1-\alpha} \, ,
\nonumber
\end{equation}
which on simplification results in (\ref{eq:astep3}).
This completes the proof of $Proposition$ 2. \hfill \rule{1.5ex}{1.5ex}

The result presented in (\ref{eq:astep3}) is valid for
any $\alpha$-Fair scheduler.
Further, it is observed from (\ref{eq:astep3}) that the optimal time fraction for CoMP users $\theta_{k}^{*}$ depends on the set of all users $\mathcal{U}_{k}$ in the virtual cluster $k$ irrespective of whether it is CoMP or non-CoMP.
The optimal CoMP time fraction $\theta_k^*$
for some commonly used $\alpha$-Fair schedulers
is presented in Table~\ref{tab:2}.
Note that for a proportional fair scheduler ($\alpha=1$),
$\theta_k^*$ is independent of
the user link rates and the time allocated to each user.
In Table~\ref{tab:2},
the $N_{nc}$  and $N_c$ in a virtual cluster $k$ are given by
\begin{eqnarray}
N_{nc}=\sum\limits_{s \in \mathcal{S}_{k}} \sum\limits_{u \in \mathcal{U}_{k}}(1-z_{u,s})x_{u,s}  \label{eq:n_nc} \mbox{, and}\\
N_{c}=\sum\limits_{u \in \mathcal{U}_{k}}\sum\limits_{s \in \mathcal{S}_{k}}z_{u,s}x_{u,s} \label{eq:n_c} \mbox{, respectively.}
\end{eqnarray}
Next, we present
a re-framed and simplified
BSS with CoMP optimization problem
for the center cluster $q$.

\section{BSS with CoMP}
The simplified problem of BSS with CoMP for a given $\Gamma_{d}$, and the optimal $\beta_{u,s}$, $\beta_{u,k}$, and $\theta_{k}$
obtained from 
(\ref{eq:aa1}), (\ref{eq:aa2}), and (\ref{eq:astep3}), respectively,
is formulated as follows
\begin{eqnarray}
\mathbb{B}^{*}  &:& \ \max_{w_{b}}\sum\limits_{b \in \mathcal{B}_{q}} w_{b} \label{eq:cb_12} \\
\mbox{s.t.} & & (\ref{eq:21}) \, ,(\ref{eq:31}) \, , (\ref{gmod1}) \, , (\ref{gmod2}) \, , (\ref{eq:15}) \, , (\ref{eq: bin_comp1})\nonumber \\
\lambda_{u} & = & \Big[ \sum\limits_{k \in \mathcal{K}_{q}} \sum\limits_{s \in \mathcal{S}_{k}}(1-\theta_{k}^*)x_{u,s}(1-z_{u,s})\beta_{u,s}^*r_{u,s} \ + \  \nonumber \\ & &
\sum\limits_{k \in \mathcal{K}_{q}} \sum\limits_{s \in \mathcal{S}_{k}} \theta_{k}^*x_{u,s}z_{u,s}\beta_{u,k}^*r_{u,k} \Big] > R \  \ \forall u \in \mathcal{V}_{q} \, ,\label{eq:c_17} \\
%x_{u,s}& =& \begin{cases}
%		1, & \text{if $s = \arg \max_s\{\gamma_{u,s}^m$\}},\\
%		0, & \text{otherwise},
%	 \forall u \in \mathcal{V}_q, \forall s \in \mathcal{W}_q ,\end{cases} \label{eq:15} \\
%z_{u,s}&=&\begin{cases}
%		1, & \text{if $\gamma_{u,s}^m\leq \Gamma_{d}x_{u,s},  \ s \in \mathcal{S}_{k}, \ k \in \mathcal{K}_{q} \  s.t. \  |\mathcal{S}_{k}| > 1$}.\\
%		0, & \text{otherwise}, \ \forall u \in \mathcal{V}_q, \forall s \in \mathcal{W}_q,	\end{cases}  \label{eq: bin_comp1} \\
\beta_{u,s}^* & &\mbox{is as in (\ref{eq:aa1}),}  \, \forall u \in \mathcal{V}_{q}, \ \forall s  \in \mathcal{W}_{q} \, , \label{eq:beta_b1} \\
\beta_{u,k}^* & &\mbox{is as in (\ref{eq:aa2}),}  \ \forall k \in \mathcal{K}_{q}  \, , \label{eq:beta_b2} \\
\theta_{k}^* & &\mbox{is as in (\ref{eq:astep3}),} \, \ \forall k \in \mathcal{K}_{q} \, ,
	\label{eq: theta_range1} 
\end{eqnarray}
where, the objective function in (\ref{eq:cb_12})
is the same as in (\ref{eq:12}),
the constraints (\ref{eq:21}), (\ref{eq:31}), (\ref{gmod1})--(\ref{eq: bin_comp1})
are required as in (\ref{eq:12}).
However, (\ref{eq:c_17}) which
is the resultant rate of a user
with BSS and CoMP
is now computed using
$\beta_{u,s}^*$, $\beta_{u,k}^*$,
and $\theta_{k}^*$
from (\ref{eq:beta_b1}), (\ref{eq:beta_b2}), and (\ref{eq: theta_range1}), 
that are obtained using
(\ref{eq:aa1}), (\ref{eq:aa2}), and
(\ref{eq:astep3}), respectively.
Note that although the optimization problem
presented in (\ref{eq:cb_12}) is relatively
simpler than (\ref{eq:12}),
it is still an MINLP.
Hence, we next present a heuristic that solves
the BSS with CoMP optimization problem.

\begin{algorithm}[t]
\begin{algorithmic}[1]
\STATE{INPUTS  : $\{P_{s}^m h_{u,s}^m\}$, $\mathcal{V}_q$, $\Gamma_{d}$, $R$}, $\{\mathcal{Z}_{a1/a2}^{j}\}$
\STATE{OUTPUTS : $\{\lambda_{u}\}$}, $\mathcal{Z}_{a1/a2}^{*}$
\STATE{Sort $\mathcal{Z}_{a1/a2}$ in increasing order of energy consumption}
\STATE{Initialize : $J=|\{\mathcal{Z}_{a1/a2}^{j}\}|$, j=1}
\STATE{\textbf{Repeat}}
\STATE{Initialize : u=1, $\{z_{u,s}\}=0$}
\STATE{\textbf{Repeat}}
\STATE{Sort $\{P_{s}^m h_{u,s}^m\}$ in decreasing order
and set $x_{u,s}=1$}
\STATE{\ \ \ $\gamma_{u,s} = f(\{P_{s}^m h_{u,s}^m\})$ as in (\ref{gmod1})}
\IF{$\gamma_{u,s} \leq \Gamma_{d}$}
\STATE{\ \ \ $\gamma_{u,k} = f(\{P_{s}^m h_{u,s}^m\})$ as in (\ref{gmod2})}       
       \STATE{$z_{u,s}=1$}
\ELSE
       \STATE{$z_{u,s}=0$}       
\ENDIF
\STATE{\ \ \ Set $u=u+1$}
\STATE{\textbf{Until} $u \geq |\mathcal{V}_q|+1$}
\STATE{Set u=1}
\STATE{\textbf{Repeat}}
\STATE{ \ \ \ Compute $\lambda_{u}$ as in (\ref{eq:17})}
\STATE{\ \ \ Set $u=u+1$}
\STATE{\textbf{Until} $u \geq \mathcal{V}_q +1$}
\IF{$\min \{\lambda_{u}\} < R$ and $j < J$ }
\STATE{$j=j+1$}
\STATE{\textbf{Goto} Step. $6$}
\ELSE
\STATE{$\mathcal{Z}_{a1/a2}^{*} = \mathcal{Z}_{a1/a2}^{j}$}
\STATE{\textbf{Goto} Step. $31$}
\ENDIF
\STATE{\textbf{Until} $j > J$ }
\STATE{\textbf{Stop}}
\end{algorithmic}
    \caption{Dynamic Base Station Switching with CoMP}\label{algo:bss_al}
\end{algorithm}

\section{Proposed Heuristic for BSS with CoMP}

In this section,
we present a heuristic
that selects the optimum
BSS pattern for a pre-determined set of virtual clusters
that perform CoMP in the center cluster $q$.
The proposed heuristic assumes
that the set of users $\mathcal{V}_q$ and the set
of received powers for any user $u$ from any sector $s$,
represented by $\{P_{s}^m h_{u,s}^m\}$
is available.
The heuristic considers
a set of BSS patterns
denoted by $\{\mathcal{Z}_{a1/a2}^{j}\}$.
Note that any element $\mathcal{Z}_{a1/a2}^{j}$ of this set
is equivalent to a unique combination of $\{w_b\}$,
the binary BSS indicator
variables specified in (\ref{eq:31}).
The heuristic also takes $\Gamma_{d}$ and $R$
as an input.
The set of BSS patterns is first sorted in an increasing
of energy consumption
such that any BSS pattern $\{\mathcal{Z}_{a1/a2}^{j}\}$ consumes
less than equal to the energy consumed by $\{\mathcal{Z}_{a1/a2}^{j+1}\}$.
The heuristic starts with
least energy consuming BSS pattern.
Next, the set of
received powers $\{P_{s}^m h_{u,s}^m\}$ is sorted
for any user $u$ from all sectors $s$.
Using this operation for every user $u$,
the sector $s$ from which it receives maximum power
is identified and $x_{u,s}$ is set as 1.
Next, given $R$
it is decided whether a user $u$
is a CoMP or a non-CoMP user.
Then,
for the BSS pattern under consideration,
the received SINRs from the corresponding sector
or virtual cluster is computed using
(\ref{gmod1}) or (\ref{gmod2}), respectively.
Note that 
(\ref{gmod1}) and (\ref{gmod2})
consider only the BSs that are still in
ON state for the SINR calculations.
In a separate loop over the number of users, i.e.,
$|\mathcal{V}_q|$, the rate of each user is computed.
This is required as the user association and
SINRs are used to compute the rate of
all users in the system as in (\ref{eq:17}).
In case all users receive a rate higher than
the rate threshold $R$
then the heuristic stops and
selects this BSS pattern as the optimum pattern.
Otherwise,
the number of switched on BSs is
increased and the described steps
are repeated for the next BSS pattern.
The heuristic runs till either a optimum BSS
pattern is obtained or all BSs are in ON state.
The heuristic is presented
as a pseudo-code in Algo.~1. 

The practical implementation of the proposed heuristic will run
at any one of the BSs in a cluster $q$, such
that this particular BS acts as a centralized controller and takes the decisions
for all the BSs in the cluster.
Given a user realization,
the centralized controller
decides whether CoMP should be performed or not based on the operator's rate threshold, and CoMP SINR threshold.
The user's information particularly SINR
and rate has to be sent to the centralized controller
so that it can decide
the CoMP configuration,
CoMP time fraction, and
user scheduling time fractions.
This will result in additional
overhead on the backhaul
which can be compensated in terms of improvement in coverage and energy savings.
The computational complexity of the proposed heuristic
for every user realization is
$O(J ( |\mathcal{V}_q| |\mathcal{B}_q| +|\mathcal{V}_q|))$.
Note that worst case $J$ is equal to $2^{|\mathcal{B}_q|}$.
However, in practice, operators can optimize
and choose from a lower number of BSS patterns.
For example, in the numerical results
presented next,
we consider $J$ equal to five BSS patterns.

\begin{table}[t]
	\begin{center}
		\caption{Simulation Parameters}	
		\label{tab:sim_set}
		\begin{tabular}{|c|     c      |} 
			\hline
			$B$ & 49 \\
			\hline
			Inter-site Distance     &  500 m      \\ %[0.5ex] 
			\hline
			Penetration loss ($\upsilon$)  &  20 dB       \\ %[0.5ex] 
			\hline
			Loss due to Log-normal&  \\ shadowing ($\rho$)   & Standard deviation of 8 dB \\ 
			\hline  
			P$_{BS}$ & 46 dBm\\
			\hline			
			$\sigma^2$ & 2.2661e-15\\
			\hline			
			PL(d) & 136.8245+(39.086(log$_{10}$d-3)) \cite{r20} \\			
			\hline
			$M$ & 99 \\			
			\hline
			Subchannel Bandwidth & 180 KHz \\
		    \hline
		    SC$_{OFDM}$ & 12\\
		    \hline
		    SY$_{OFDM}$ & 14\\
		    \hline
		    T$_{Subframe}$ & 1 ms \\ 
		    \hline
		    Cluster size & 7\\		   		    
			\hline						
		\end{tabular}
	\end{center}	
\end{table}

\section{Numerical Results}
We consider a center cluster with 7 BSs.
To model the interference suitably,
we consider a wrap-around system with 6 clusters
of 7 BS each around the center cluster.
We consider the simulation parameters specified by 3GPP
for an urban homogeneous cellular environment
as given in \cite{r20}.
Thus, a total of 49 BSs are considered for simulations
with inter-site distance of 500 m.
The users are distributed uniformly randomly with the
appropriate user density ($\mu$) over the entire simulations area.
We consider 500 user location realizations.
For each location realization the results
are averaged over 50 independent fading realizations.
The simulation parameter details
are given in Table~\ref{tab:sim_set}.
To study the impact of change in $\mu$
over the system performance,
we vary the average user density from $20$ to
$160$ users per $km^2$.

\begin{figure}[t]
	\begin{center}	
		\centering
		\includegraphics[height=2.3in,width=\columnwidth]{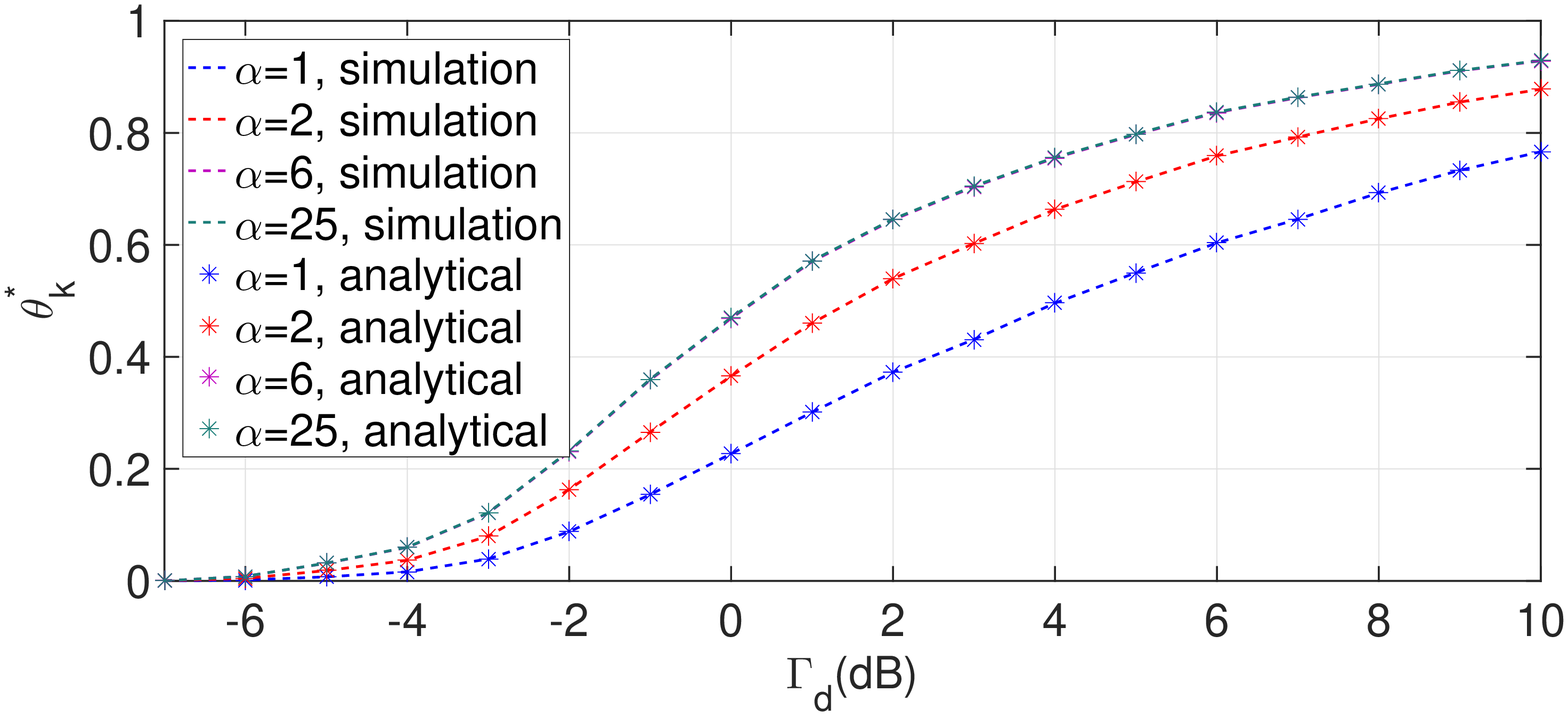}			
		\caption{Variation of optimal CoMP time fraction
		($\theta_{k}^*$) with respect to CoMP SINR threshold
		($\Gamma_{d}$)
		for various fairness parameter ($\alpha$).}
		\label{fig:theta}
	\end{center}
\end{figure} 

The variation of $\theta_k^*$ with respect to
$\Gamma_{d}$ is shown in Fig.~\ref{fig:theta}
for various values of $\alpha$.
Note that the optimal value of $\theta_k$
obtained via exhaustive search in simulations
matches with the $\theta_k^*$ derived in (\ref{eq:astep3}).
Further, the optimal CoMP time fraction
increases with an increase in
the CoMP SINR threshold as more
number of users become CoMP users
with increase in $\Gamma_{d}$.
The increase in $\alpha$ values makes the
$\alpha$-Fair scheduler allocate more
resources to edge users.
Hence, an increase in the fairness parameter
$\alpha$ results in an increase in $\theta_k^*$
for the same value of $\Gamma_{d}$.
The increased $\theta_k^*$ ensures
that the edge users (with SINR $\leq \Gamma_{d}$)
will be served as CoMP users
and receive more downlink time fraction.

\begin{figure}[t]
	\begin{center}	
		\centering
		\includegraphics[height=2.3in,width=\columnwidth]{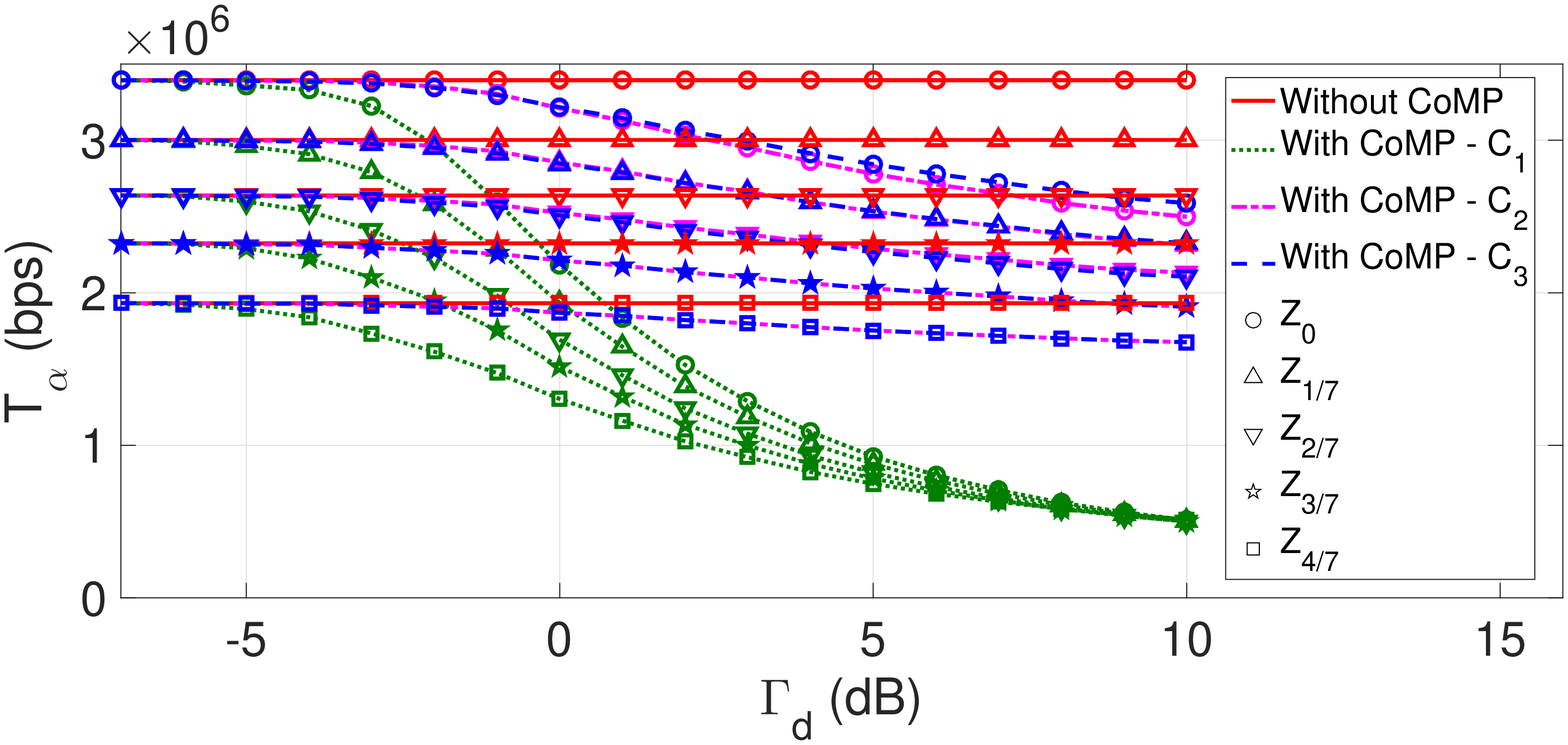}			
		\caption{
Variation of system throughput ($T_\alpha$) with respect to CoMP SINR threshold ($\Gamma_{d}$), with and without CoMP, for various BSS patterns.}
		\label{fig:tput_4bss}
	\end{center}
\end{figure} 

\begin{figure}[t]
	\begin{center}	
		\centering		\includegraphics[height=2.3in,width=\columnwidth]{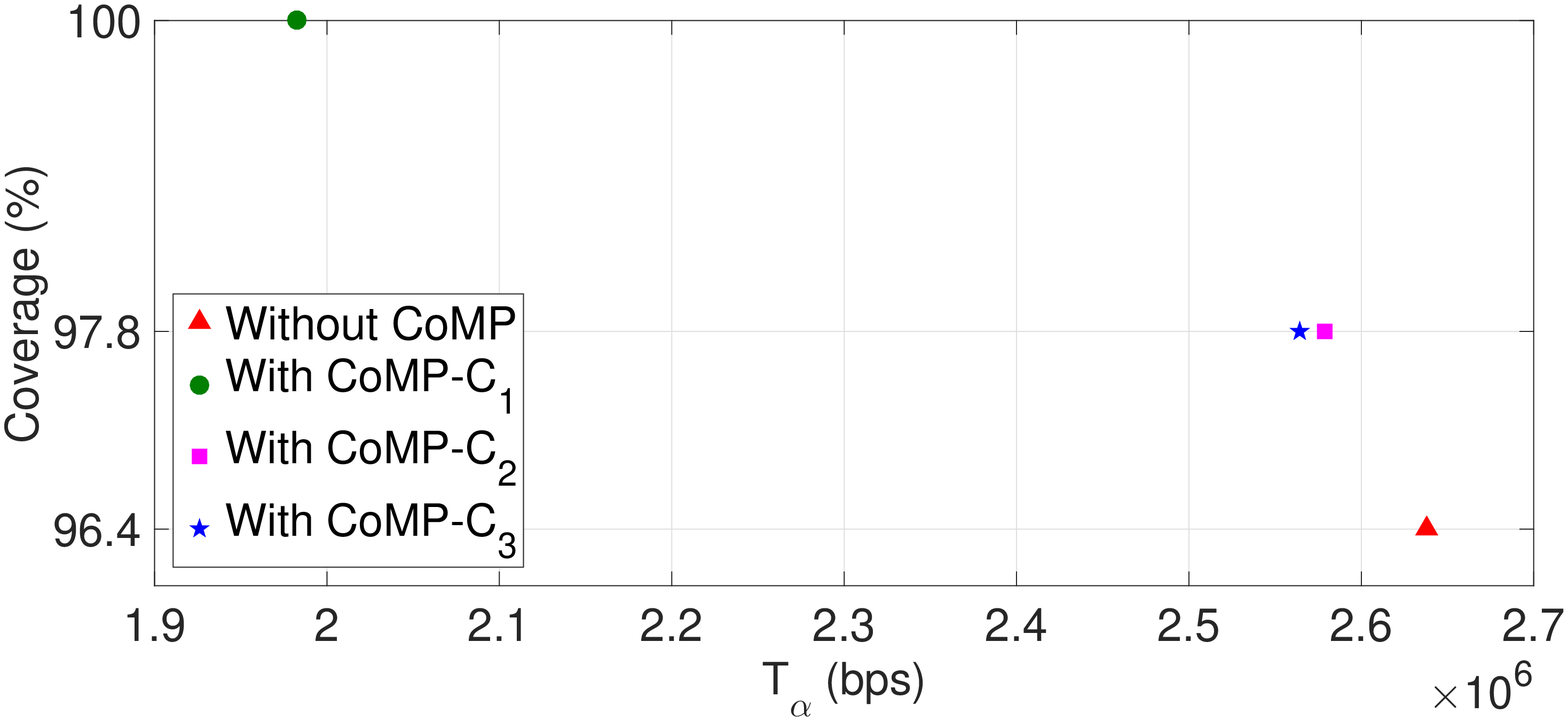}			
		\caption{Coverage and throughput trade-off for user density of $60/km^2$ and BSS pattern $\mathcal{Z}_{3/7}$.}
		\label{fig:ct}
	\end{center}
\end{figure} 

The throughput metric
corresponding to a $\alpha-$Fair scheduler is given in (\ref{eq:tput_eqn}).
The variation of $T_\alpha$ with respect to
$\Gamma_{d}$, different BSS patterns,
and $\alpha=1$ is presented in Fig.~\ref{fig:tput_4bss}.
Note that the throughput decreases as more BSs
are switched off.
Further, even with various BSS patterns,
the without CoMP scenario, CoMP configuration
$C_3$, $C_2$, and $C_1$ are in decreasing order
of throughput.
This is due to the rate and coverage
trade-off between these configurations.
To better illustrate this,
we present the rate and coverage
trade-off for the BSS pattern
$\mathcal{Z}_{3/7}$
in Fig.~\ref{fig:ct} for the user density of $60$ users/km$^{2}$.
The probability of coverage is as defined in Section IIF for SINR coverage.
Note that an operator can run the network
without CoMP for maximum throughput at the cost of coverage.
On the other hand,
all sectors CoMP in $C_1$ can provide
maximum coverage at the cost of throughput.

\begin{figure}[t]
	\begin{center}	
		\centering		\includegraphics[height=2.3in,width=\columnwidth]{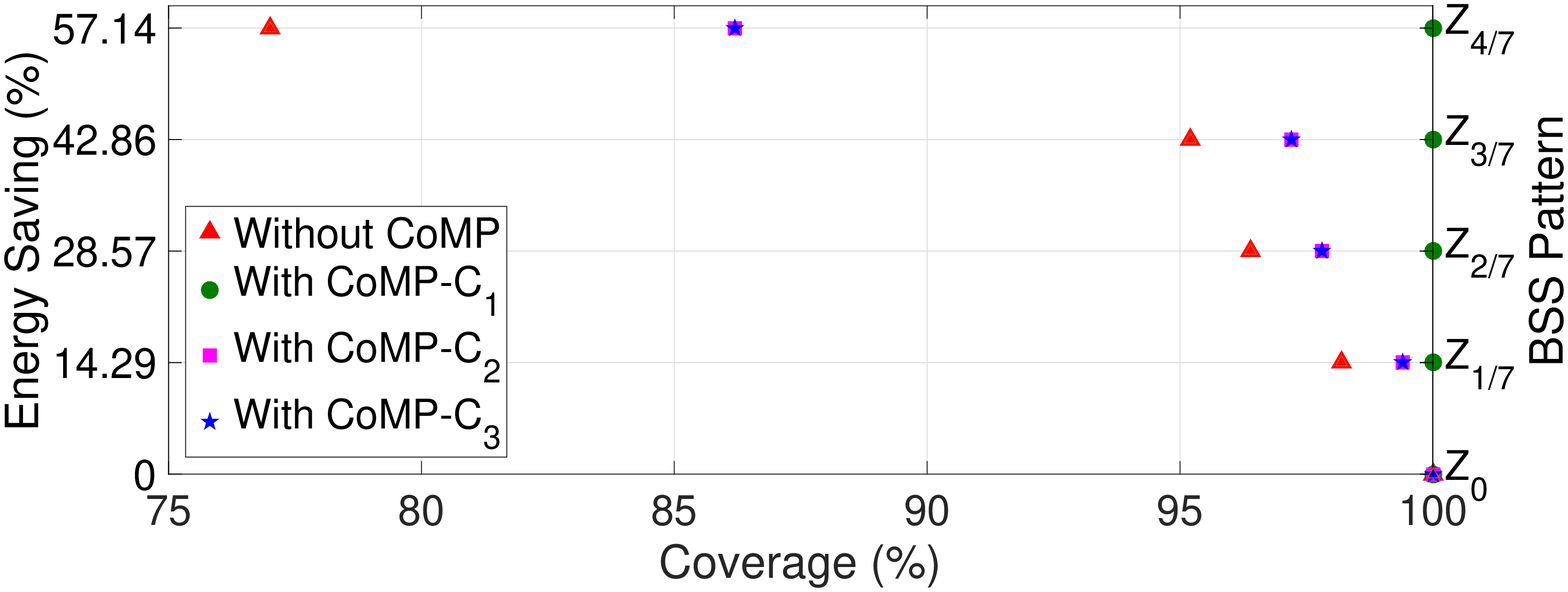}			
		\caption{Energy and coverage
		trade-off for user density of $60/km^2$, various BSS patterns, and CoMP configurations (Note that the corresponding throughput is depicted in Fig.~\ref{fig:et}).}
		\label{fig:ec}
	\end{center}
\end{figure}

The trade-off between
percentage energy savings
and 
coverage is presented for $\mathcal{Z}_0$ (all BSs in ON state)
and BSS patterns  $\mathcal{Z}_{1/7}$, $\mathcal{Z}_{2/7}$, $\mathcal{Z}_{3/7}$ and $\mathcal{Z}_{4/7}$, and
various modes of CoMP operations in Fig.~\ref{fig:ec}.
The results considered are
for BSS patterns shown in Fig.~\ref{fig:z2}, ~\ref{fig:z4}, ~\ref{fig:z8}, and \ref{fig:z9}.
An increase in
the number of switched off BSs
results in decrease in the coverage probability
for any particular CoMP configuration.
However, switching off BSs
also increases the 
percentage energy savings.
Thus, an operator can use the results
in \ref{fig:ec} to
select the appropriate point of operation
and the corresponding trade-off between
percentage energy savings
and 
coverage.

\begin{figure}[t]
	\begin{center}	
		\centering		\includegraphics[height=2.3in,width=\columnwidth]{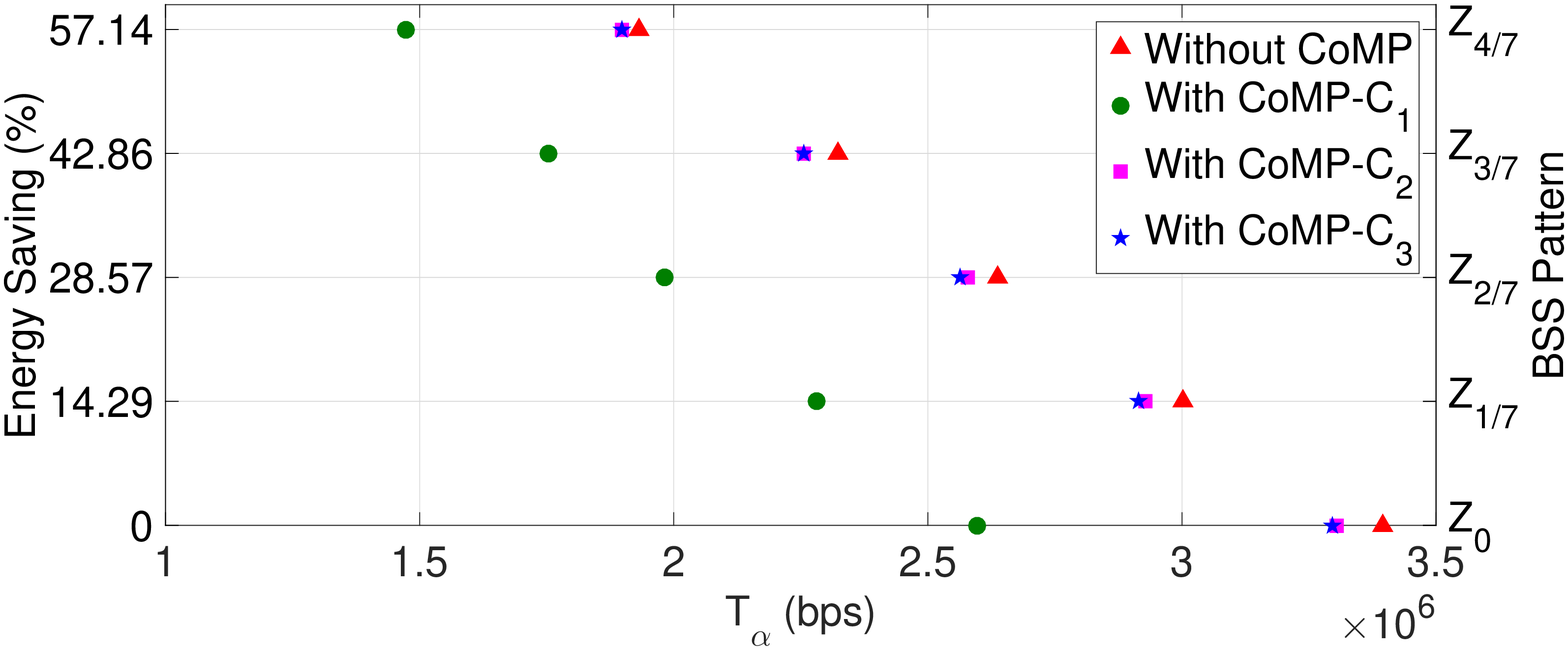}			
		\caption{Energy and throughput trade-off for user density of $60/km^2$, various BSS patterns, and CoMP configurations.}
		\label{fig:et}
	\end{center}
\end{figure}

The corresponding energy and throughput
trade-off for the various
BSS patterns and CoMP modes in Fig.~\ref{fig:ec}
is presented in 
Fig.~\ref{fig:et}.
Note that an operator should jointly utilize
the Fig.~\ref{fig:ec} and Fig.~\ref{fig:et}.
For example,
in Fig.~\ref{fig:ec}, the coverage probability
of $C_1$ is higher than $C_2$ for all BSS scenarios.
Whereas, in Fig.~\ref{fig:et},
the throughput of $C_1$ is lower
than $C_2$ for all BSS scenarios.
Thus, multiple configurations of
BSS with CoMP can be used to achieve
various trade-offs between energy, coverage
and rate trade-off
which a traditional without CoMP system
does not offer.

\begin{figure}[t]
	\begin{center}	
		\centering
		\includegraphics[height=2.3in,width=\columnwidth]{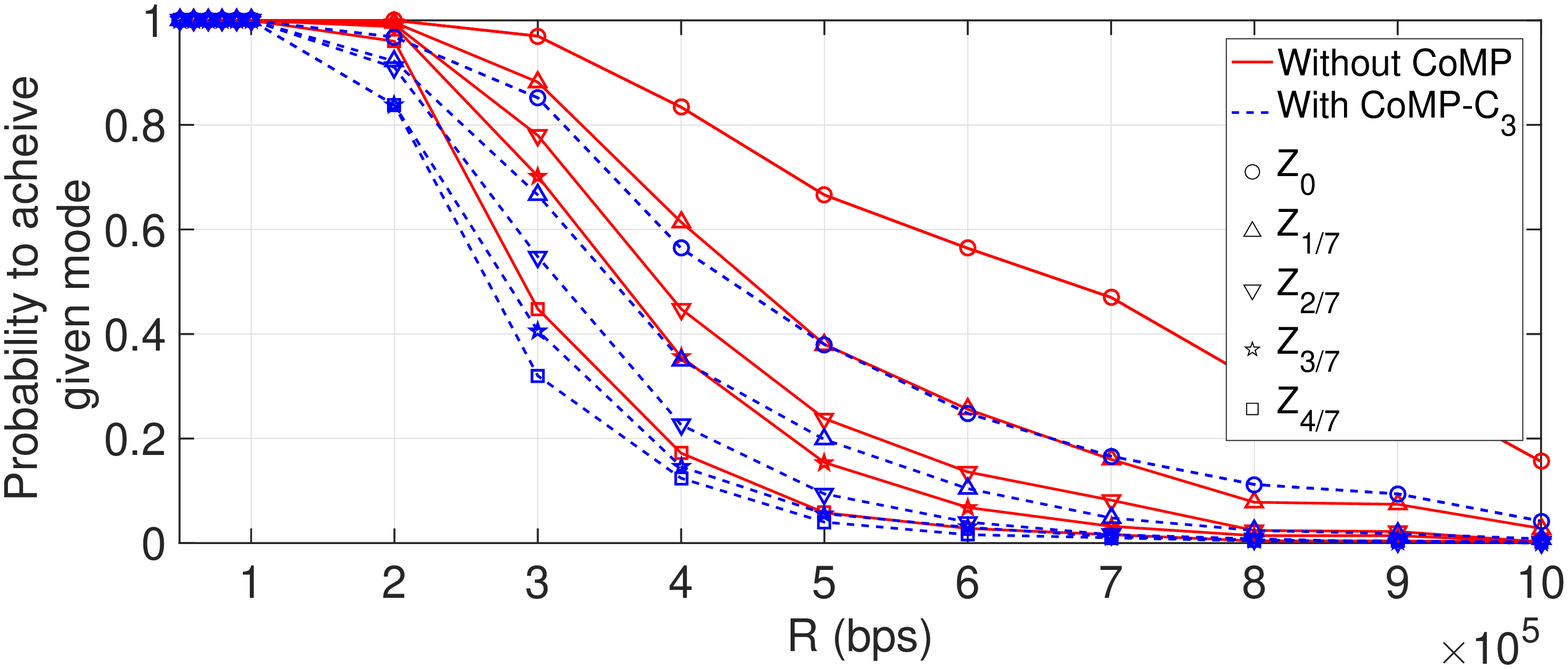}			
		\caption{
Variation of rate coverage with respect to rate threshold
($R$), for various BSS patterns in configuration $C_{3}$, $\alpha=1$, and $\Gamma_{d} = -1$ dB.}
		\label{fig:rcov_5modes}
	\end{center}
\end{figure}

\begin{figure}[t]
	\begin{center}	
		\centering
		\includegraphics[height=2.3in,width=\columnwidth]{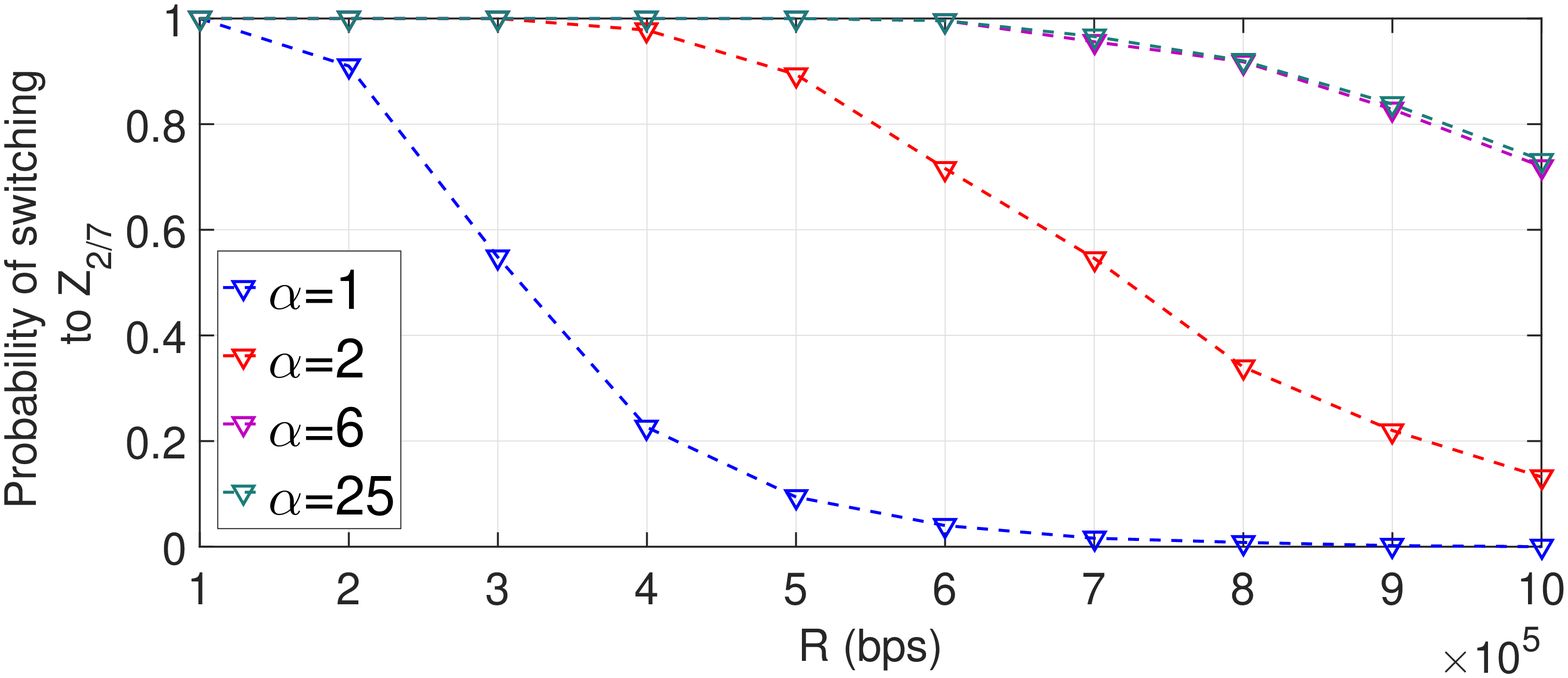}			
		\caption{Variation of rate coverage with respect to rate threshold ($R$), for various $\alpha$, BSS pattern $\mathcal{Z}_{2/7}$ in configuration $C_{3}$,
		and $\Gamma_{d} = -1$ dB.}
		\label{fig:rcov_alp}
	\end{center}
\end{figure}

For the next two set of results,
we focus on $C_3$
as it results in least loss in throughput
in comparison to without CoMP scenario.
The rate coverage as defined in 
Section IIF is presented in
Fig.~\ref{fig:rcov_5modes} and Fig.~\ref{fig:rcov_alp}
for $\alpha=1$ and $\Gamma_{d} =-1dB$ .
In Fig.~\ref{fig:rcov_5modes},
the probability to operate in
with a BSS pattern while ensuring
the user rates to be higher than
the rate threshold $R$
is presented
for without CoMP and
with CoMP configuration $C_3$.
The Fig.~\ref{fig:rcov_5modes}
shows that
to maintain the same rate coverage
with larges energy savings the 
system has to reduce the rate threshold $R$.
Further,
for the same $R$,
BSS patterns with higher energy savings are less
probable.
Note that Fig.~\ref{fig:rcov_alp} is for
BSS pattern $\mathcal{Z}_{2/7}$.
It is observed from
Fig.~\ref{fig:rcov_alp} that
the probability for selecting the BSS pattern
increases with increase in $\alpha$.
Thus, Fig.~\ref{fig:rcov_5modes} and Fig.~\ref{fig:rcov_alp} also depict
the rate-coverage and energy
trade-off discussed earlier
from a probabilistic perspective.

\begin{figure}[t]
	\begin{center}	
		\centering
		\includegraphics[height=3in,width=\columnwidth]{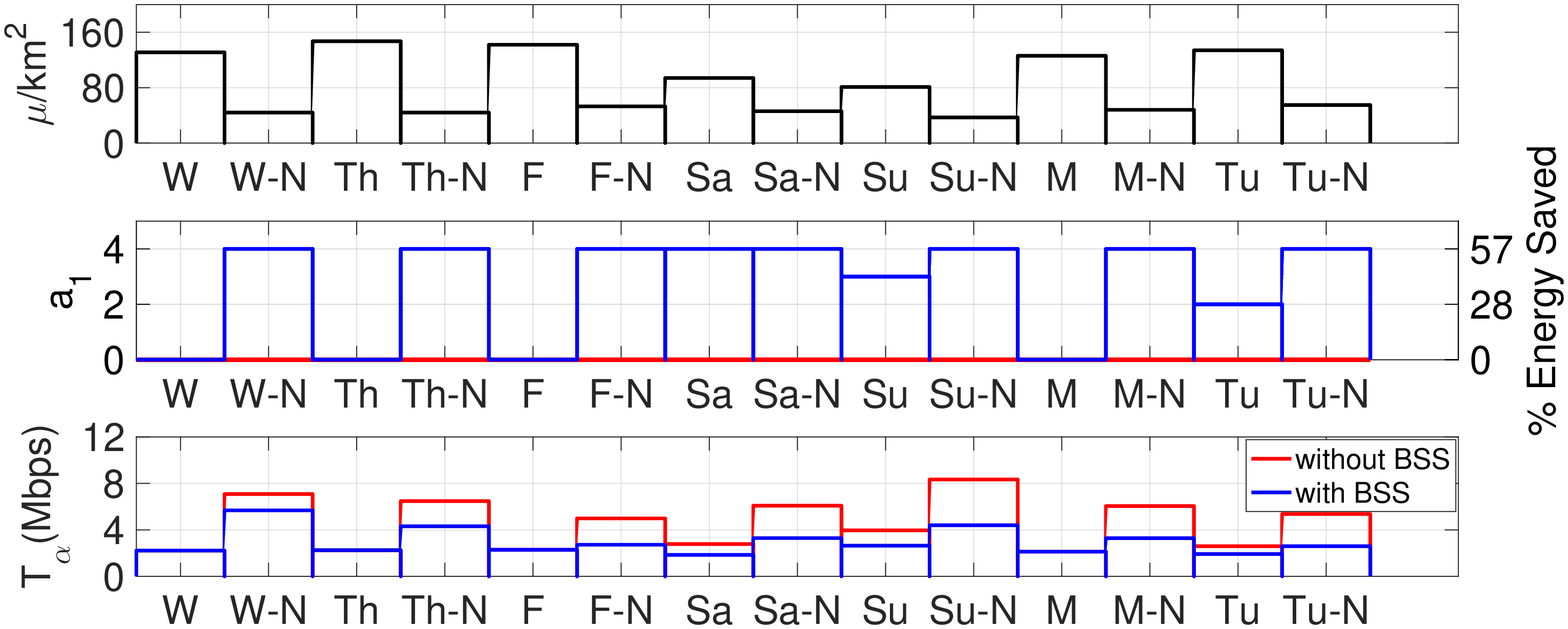}			
		\caption{Performance of the proposed heuristic in a varying traffic scenario when $R = 0.2Mbps$, a1 is number of switched of BSs, $\mu$ is user density.}
		\label{fig:algo_result}
	\end{center}
\end{figure}

In Fig.~\ref{fig:algo_result},
the result from the heuristic proposed in Section VI
is presented.
We select $R$ as $0.2$ Mbps.
A snapshot of traffic profile variation is selected and a optimum BSS pattern ($\mathcal{Z}_{a1/a2}^{j}$) is selected based on the given operator rate threshold $R$.
In Fig.~\ref{fig:algo_result}, a1 represents the number of BSs switched off and correspondingly the
percentage energy saved.
It is observed from Fig.~\ref{fig:algo_result}
that there is some decrease in overall
throughput whenever BSs are switched off.
However, the loss in throughput
is accompanied with significant gain in
terms of energy savings.
Thus, the proposed heuristic
ensures maximum energy savings,
without loss in coverage,
at the cost of high rate users.

%\begin{figure}
%\begin{algorithmic}[1]
%
%\Repeat
%\Comment{forever}
%\State this\Until{condition satisfired}
%\Statex
%\end{algorithmic}
%\end{figure}

\section{Conclusion}

We have shown that loss in SINR coverage
due to BSS can be compensated by CoMP transmission.
We have formulated the joint
BSS and CoMP problem as an optimization
problem.
The optimal solutions for a decomposed
CoMP resource allocation and
user scheduling problem have been derived.
The derived results hold for arbitrary BSS patterns, and given a BSS pattern
can also be applied to any cluster.
The derived results have been used to formulate
a simplified BSS with CoMP problem.
A heuristic has been presented that solves
the BSS with CoMP problem dynamically.
Through numerical results it has been shown
that the derived results match closely with
simulations.
Further,
we have shown that BSS with CoMP
can be used to achieve various possible
trade-offs in energy savings, coverage, and throughput.
In future, the presented work
will be extended using
a stochastic geometry based framework
for arbitrary cluster size.

\bibliographystyle{ieeetr}
\bibliography{manasa}

% As a general rule, do not put math, special symbols or citations
% in the abstract

\end{document}